\documentclass[%
 reprint,
superscriptaddress,
showpacs,preprintnumbers,
 amsmath,amssymb,
 aps,
pra,
]{revtex4-1}
\usepackage{xcolor}
\usepackage{times}
\usepackage{graphicx}
\usepackage{dcolumn}
\usepackage{bm}
\usepackage{epstopdf}
\usepackage{hyperref}
\newcommand*\chem[1]{\ensuremath{\mathrm{#1}}}

\begin{document}

\title{\textbf{Emergent superconductivity upon disordering a charge density wave ground state}}

\author{Anurag Banerjee}
\affiliation{Indian Institute of Science Education and Research-Kolkata, Mohanpur Campus, India-741252}

\author{Arti Garg}
\affiliation{Condensed Matter Physics Division, Saha Institute of Nuclear Physics, 1/AF Bidhannagar, Kolkata 700 064, India}

\author{Amit Ghosal}
\affiliation{Indian Institute of Science Education and Research-Kolkata, Mohanpur Campus, India-741252}

\begin{abstract}
We explore the interplay of a charge density wave (CDW) order and s-wave superconductivity (sSC) in a disordered system. 
Recent experiments on 1T-\chem{TiSe_2}, where the pristine sample has a commensurate CDW order and the superconductivity appears upon copper intercalation, motivates our study. Starting with an extended Hubbard model, with parameters which yield a CDW ground state within Hartree-Fock-Bogoliubov formalism in pure systems, we show that the addition of disorder quickly wipes out the global charge order by disrupting periodic modulation of density at some (low) strength of disorder. Along with this, the subdominant superconducting order emerges in regions that spatially anti-correlates with islands of strong local CDW order. The short-range density modulations, however, continue to persist and show discernible effects
up to a larger disorder strength. The local CDW puddles reduce in size with increasing disorder and they finally lose their relevance in effecting the properties of the system. Our results have strong implications for the experimental phase diagram of transition metal dichalcogenides. 
\end{abstract}

\maketitle

\section{Introduction}

One common feature of nearly all unconventional superconductors is that the superconductivity coexists and competes with other symmetry broken states in a wide range of parameter space. These include the charge orders~\cite{Wise2008,Chang2012,CupratesPRB,DisorderCDWHTSc,silvaNeto2014,Croft2014} and antiferromagnetism~\cite{Sachdev1510,Lu2014,So5RMP} in cuprtate superconductors, spin density waves in pnictides~\cite{ThSDWPnictides,Wang200,SchmiedtPair,SiRev}, orbital order in Ruthenates~\cite{Ruth0,Ruth1,Ruth2}, among others. Often, the symmetry unbroken state is accessed via these competing ordered states upon destruction of superconductivity. For example, a high-$T_c$ cuprate superconductor for low to optimal doping, goes through a $d$-wave superconducting, charge ordered and pseudo-gapped phases before showing metallicity upon increasing temperatures ~\cite{Chang2012}, albeit the metal being quite unconventional~\cite{Dagotto}. Similarly in underdoped pnictides, transition from superconducting state to the metallic state occurs via an intervening AFM ordered state~\cite{Wang200}. So abundant are such examples that a scenario of competing orders has become perhaps more of a paradigm for unconventional superconductors than the metallic BCS superconductors~\cite{ThUnconvSC,Norman196}.

However, even the conventional BCS s-wave superconductors (sSC) have a competing cousin -- a charge density wave (CDW) state. In fact, the attractive Hubbard model, a minimal description for sSC~\cite{PhaseAHM}, supports a ground state which is degenerate in superconductivity and charge density wave channels~\cite{PhaseAHM,Moreo,Miller} at half-filling. On the other hand, it is established that such a commensurate CDW order is fragile to perturbations -- it can be destroyed easily by doping away from half-filling~\cite{Moreo} or by introducing disorder~\cite{RTS} or by allowing long range hopping~\cite{DosSantos}. It is this disorder dependence  of the CDW order that remains in our prime focus in the present study. Disorder driven inhomogeneities are known to generate intriguing local properties in s-wave superconductors~\cite{GTR}. The question that is addressed in this paper is: Are there signatures of short-ranged CDW order which alters the properties of an underlying disordered s-wave superconductor?

While the interplay of sSC and CDW had been attracting research for a long time~\cite{CDWoldRev,Mutka83,McMillan0}, it has come to the forefront of renewed interest in condensed matter physics with the discovery of superconductivity in transition metal dichalcogenides (TMD). For example, superconductivity emerges in 1T-\chem{TiSe_2} upon copper intercalation~\cite{Cava0} (with doping fraction $x$). Note that the intercalation introduces stoichiometric disorder in these materials, in addition to changing the density of the charge carriers. Most TMDs support a commensurate charge density waves order in their ground state. In case of pristine 1T-\chem{TiSe_2} (i.e. $x=0$), such a commensurate CDW order sets in below the transition temperature, $T_c \sim 200K$. With \chem{Cu} intercalation, $T_c$ degrades quickly, and conventional superconductivity arises for doping level $x \geq 0.04$~\cite{Cava0}, reaching its maximum strength at $x \sim 0.08$, with $T_c \sim 4.2K$. With further increase of \chem{Cu} intercalation, $T_c$ keeps decreasing forming a superconducting dome in $(T,x)$-phase diagram, reminiscent of cuprate superconductivity~\cite{CVarma,Dagotto}. Surprisingly, the initial rapid decay of CDW order slows down considerably beyond dopings which marks the onset of superconductivity, and an incommensurate charge density wave (ICDW) order survives beyond this point. Such ICDW phase is found to persist along with superconductivity at doping as large as $x \approx 0.11$~\cite{Abbamonte,Samuelymagnetic}.

A closer look into the regions of coexistence using the scanning tunneling microscope (STM) reveals~\cite{YanSTM} that the incommensuration occurs through the formation of domain walls. These separate regions of CDW order such that two adjacent CDW domains have a $\pi\text{-phase}$ shift in their charge modulation. This $\pi$-phase shift in domains is associated with a splitting of the peaks in the Fourier transform of the local conductance obtained by the STM measurements~\cite{YanSTM}. It is these domain wall regions where superconductivity is found to nucleate keeping these two independent orders (i.e., sSC and CDW) spatially apart. It is only such ICDW ordering tendencies which were found to coexist with sSC~\cite{Cava0, Abbamonte}. Naturally, such an incommensurate CDW phase lacks a long range coherence. On the other hand, coexistence of commensurate CDW with sSC are rarely found~\cite{YanSTM}.

A qualitatively similar phase diagram has been found, not only using \chem{Cu}-intercalation as the tuning parameter, but also by applying pressure on 1T-\chem{TiSe_2}~\cite{PressureTiSe2,PressurePRL}, self doping with $\chem{Ti}$~\cite{TiDop} and intercalating with \chem{Pd}~\cite{Pddop}. The formation of domains of CDW and the nucleation of superconductivity within the domain walls is not special just for 1T-\chem{TiSe_2}, but is found for several other TMDs. For example, by applying pressure in 1T-\chem{TaSe_2}~\cite{SiposTaS2}, doping with \chem{Se}~\cite{SedopTaSe2} one observes similar phenomena. Also the 2H polytype of the \chem{TaSe_2} shows similar transitions by \chem{Cu} doping~\cite{SCinCuxTaSe2}, introducing crystallographic disorder by intercalating \chem{Se}~\cite{Li2017}, and irradiation induced defects~\cite{Mutka83}. Many other TMDs like 2H-\chem{NbSe_2}~\cite{Arpes2HNbSe2}, 2H-\chem{TaSe_2} and even cuprates~\cite{DisorderCDWHTSc} follow more or less a similar trend. All these evidences point towards a paradigm that is expectedly universal and doesn't require fine tuning of parameters of these widely different materials.

\begin{figure}
      \includegraphics[width=8.5cm,height=10.0cm,keepaspectratio]{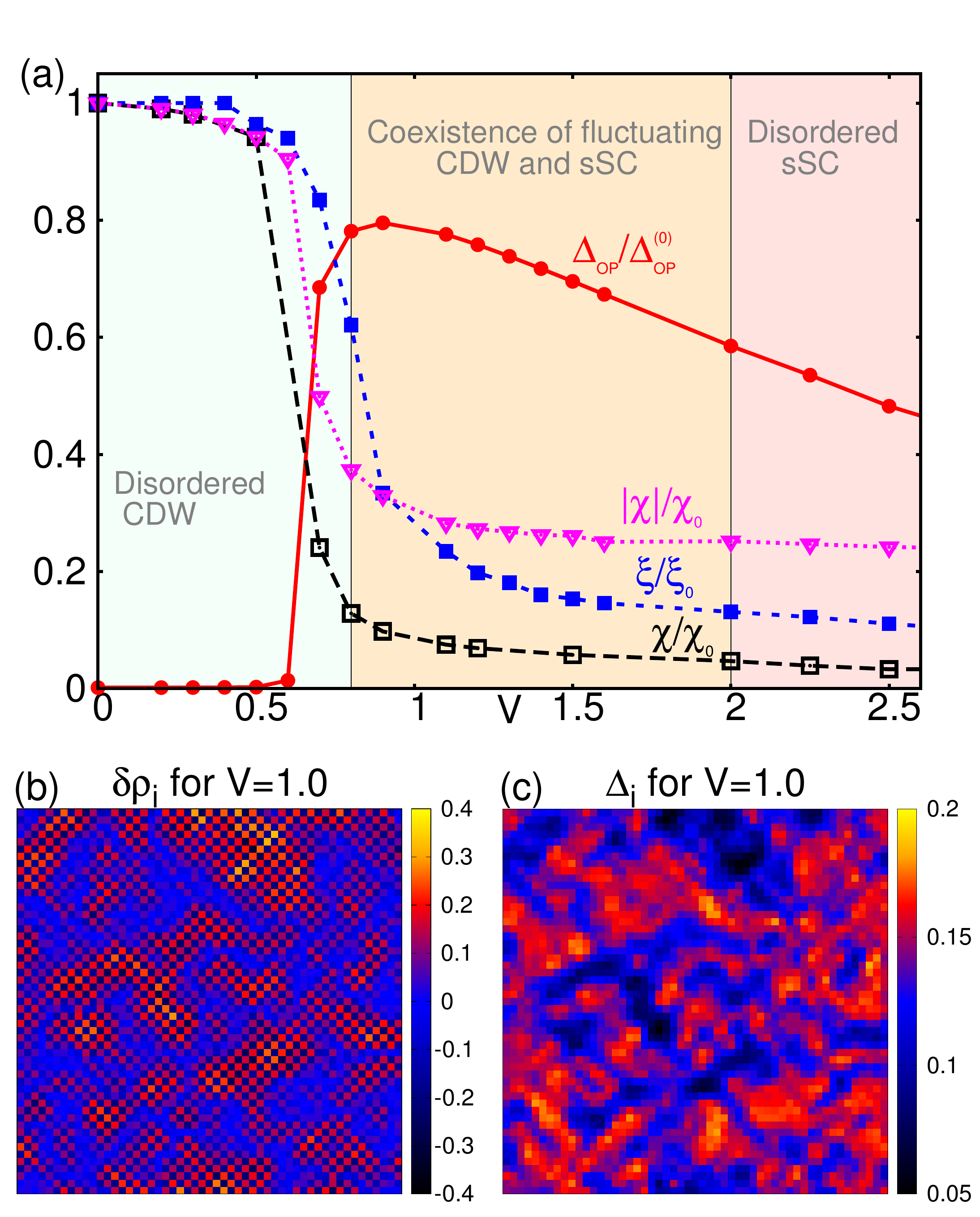}
      \caption{(Color Online) Panel \textbf{(a)} depicts the disorder ($V$) dependence of competing global orders, namely, the CDW order ($\chi$, shown in black) and superconducting order ($\Delta_{\rm OP}$, shown in red). $\chi$ vanishes at $V_1 \sim 0.8$ (shown by the first vertical line), whereas, $\Delta_{\rm OP}$ starts to develop rapidly. The average of the modulus of local CDW order $\vert \chi \vert$ (shown in magenta), decreases with $V$, but tends to saturate at a non-zero value at large $V$ hinting survival of a short ranged CDW.
The size of these `puddles' of CDW, denotes as $\xi$ (shown in blue) decreases with $V$ and attains the smallest possible value of a few lattice spacing by $V_2 \sim 2.0$ (shown by the second vertical line, and discusses in the text), implying that the presence of short ranged CDW fluctuations up to $V_2$. Physical properties in this Beige-colored region ($V_1 \leq V \leq V_2$) carry the footprints of the local CDW fluctuations on the underlying disordered superconductor. On the other hand, for $V_2 \geq 2.0$ (in the pink phase), it is only disordered superconductivity that controls the physics, where fluctuating CDW `grains' loose their relevance. The panel \textbf{(b)} depicts the fluctuation of local density ($\delta \rho_i$) at $V=1.0$. Though the global CDW order is lost by $V=1.0$, the short range phase shifted puddles of CDW (represented by black and yellow point alternating at site to site) are clearly seen. The panel \textbf{(c)} shows the local sSC pairing amplitude ($\Delta_i$) at $V=1.0$. Comparison of panels \textbf{(b)} and \textbf{(c)} demonstrates that the sSC and CDW orders are spatially separated from each other.}  
      \label{fig:intro}
   \end{figure}

We wish to capture the qualitative physics of the aforementioned phenomena within a simple theoretical framework. Our goal is to propose a model which allows a CDW ground state in the pure system, along with a sub-dominant sSC order.

We wish to mimic the role of \chem{Cu}-intercalation in experiments by introducing disorder in our calculation. Though \chem{Cu}-intercalation has the effect of changing the carrier density in addition to including stoichiometric disorder in real materials, we will fix the electronic density to a desired value while varying the disorder strength in our calculations. This will help in developing insights into the role of impurities on interplay of underlying competing orders.
Our key results are encapsuled in a $T=0$ phase diagram which is summarized in Fig.~(\ref{fig:intro}a). By increasing disorder strength $V$, the commensurate CDW order, denoted as $\chi$, degrades quickly as shown in Fig.~(\ref{fig:intro}a), disrupting the global charge ordering at strength $V_1 \sim 0.8$ (for our choice of parameters, as elaborated in Sec.~IV). The breakdown of this order is accompanied by the formation of domain walls where the charge modulations lose their phase coherence as depicted in Fig.~(\ref{fig:intro}b). However, the presence of short range CDW fluctuations is evident from the non-zero magnitude of this modulation $\vert \chi \vert$, presented in Fig.~(\ref{fig:intro}a), which persists for larger $V$. The length-scale of coherent CDW modulation, denoted as $\xi$, decreases rather gradually compared to the average order parameter and saturates roughly around $V_2 \sim 2.0$. We will show that beyond $V_2$ locally fluctuating CDW order loses relevance, i.e. they do not alter the properties of the underlying disordered sSC. Interestingly, the domain wall regions separating coherent CDW islands nucleate sSC as evident from Fig.~(\ref{fig:intro}c). The superconducting correlations, termed $\Delta_{\rm OP}$ here, grow rapidly past $V_1$ shown in Fig.~(\ref{fig:intro}a). 

The plan of the rest of the paper is as follows. In Sec.~II we introduce the model to describe the interplay of the two orders under consideration, as well as the computational method of our study which is set at $T=0$. In particular, we discuss the observables whose disorder dependence constitute the key results in Fig.~(\ref{fig:intro}). Subsequently, we discuss in Sec.~III the phase diagram in the clean system (without disorder). This helps us identify the parameter regime in which the disorder dependence would be studied in the later sections.  In Sec.~IV we present details of our results to illustrate the nature of the interplay between the CDW order and s-wave superconductivity, in the presence of disorder. We present results for the disorder dependence of various observables, such as, the structure factor, order parameters and their spatial correlations, density of states and superfluid density. We also discuss how the spatial distribution of local orders correlate with physical observables. Finally, we conclude in Sec.~V. Some details of the methods and results are included as three appendices for completeness. Through all sections, we make connections of our findings to recent experiments. 

\section{Model and Method}
We study an extended Hubbard model with on-site attraction $(U)$ and nearest neighbor repulsion $(W)$, given by:  

\begin{align}
\mathcal{H}=-t\sum_{\langle i,j \rangle, \sigma} (c^{\dagger}_{i  \sigma} c_{j \sigma} + h.c.) -U\sum_i n_{i \uparrow} n_{i \downarrow} \nonumber \\ +W \sum_{\langle i,j \rangle \sigma \sigma^{\prime}} n_{i\sigma} n_{j \sigma^{\prime}} +\sum_{i \sigma} (V_i - \mu) n_{i \sigma}
\label{EHM}
\end{align}
The first term represents the kinetic energy due to electrons hopping to nearest neighbours of a two-dimensional $(2D)$ square lattice of $N$ sites. Here the local density operator $n_{i \sigma}=c^{\dagger}_{i\sigma} c_{i\sigma}$. The disorder potential is chosen uniformly from a box distribution $V_i \in \left[ -\tfrac{V}{2},\tfrac{V}{2}\right]$, defining $V$, the width of the distribution, as the disorder strength. The average density of electrons $\rho=N^{-1}\sum_i n_{i \sigma}$ is tuned by the chemical potential $\mu$. 

We now discuss the role of the interactions in the above Hamiltonian. The on-site attraction alone generates sSC phase for any $\rho$, whereas, it also supports a CDW phase close to half-filling $(\rho=1)$. Therefore, the attractive Hubbard model (with $W=0$) at $\rho=1$ is a minimal model to study the interplay between the sSC and CDW phases~\cite{Miller}. Of these, the CDW phase is rather delicate and disappears upon perturbing away from the half-filling~\cite{Moreo} or by introducing disorder~\cite{RTS}. We thus include nearest neighbor repulsion in addition to on-site Hubbard attraction to stabilize the CDW order in clean systems. It is this $W$-term which tilts the balance between the two broken symmetry orders favoring CDW as the ground state. Because both interaction terms individually favor a CDW modulation with an ordering wave-vector $\mathbf{q}=(\pi,\pi)$ at half-filled square lattice, we consider only this particular wavevector for all our analysis. The fate of the interplay using Hamiltonian similar to Eq.~(\ref{EHM}) had been studied in the past for pure systems~\cite{MicnasRMP}, and was also extended to include disorder. Such studies, however, ignored local effects~\cite{Micnasdis,Robdis} arising from disorder, which remains in our primary focus in the current study~\footnote{Depending on parameter regime, other ordering wave-vector might lead to a better ground state energy, but we expect that the qualitative nature of the interplay between chosen orders will remain insensitive to such details.}. 

Upon mean-field decomposition of interactions of Hamiltonian ${\cal H}$ in the Hartree, Fock and Bogoliubov channels, we get:
\begin{widetext}
\begin{equation}
\mathcal{H}^{\rm CDW}_{\rm sSC}=\sum_{\langle i,j \rangle, \sigma} -\left(t+W\Gamma_{ij} \right) \left( c^{\dagger}_{i  \sigma} c_{j \sigma} + h.c. \right) + \sum_{i, \sigma} \left( V_i-\mu-\frac{U}{2}\langle n_i \rangle \right) n_{i \sigma} + \frac{W}{2}\sum_{\langle i,j \rangle \sigma} \langle n_i \rangle n_{j \sigma} +\sum_{i} \left( \Delta_{i} c^\dagger_{i\uparrow} c^\dagger_{i\downarrow} +h.c. \right)
\label{Comp}
\end{equation}
\end{widetext}
The mean fields, namely, the local density $\langle n_i \rangle$, the sSC pairing amplitude $\Delta^{*}_{i}=-U \langle c_{i\uparrow}^\dagger c_{i_\downarrow}^\dagger \rangle$ and the Fock shift, $\Gamma_{ij \sigma}=\langle c^{\dagger}_{i\sigma} c_{j\sigma} \rangle$ are evaluated self-consistently. In the absence of any magnetic order, spin rotational symmetry ensures $\Gamma_{ij \sigma}=\Gamma_{ij  \bar{\sigma}}\equiv\Gamma_{ij}$. We redefine $t_{ij}=\left(t+W\Gamma_{ij} \right)$ for simplicity of notation.

The charge density modulation is assumed to have the following form:
\begin{equation}
\langle n_i \rangle = \rho_0(i) + \chi_i e^{i\mathbf{q}.\mathbf{r_i}}
\label{densityform}
\end{equation}
here $\mathbf{q}=(\pi,\pi)$ and $\chi_i$ is the local CDW order parameter. Assuming that the role of interactions is limited only in generating the broken symmetry orders, the background local density $\rho_0(i)$ refers to a system with $W=0, U=0$. This is nothing but the inhomogeneous density of the underlying tight binding model with disorder (i.e., Anderson Model)~\footnote{We verified that inclusion of Fock shift (self-consistently) has no discernible changes to this density profile.}. 

The CDW order parameter is obtained by averaging $\chi_i$ over all sites of the lattice for a given disorder configuration and then averaging over various independent configurations of disorder $\chi=N^{-1}\langle \sum_i \chi_i\rangle_c$. Here $\langle \rangle _c$ represents configuration averaging. Fig.~(\ref{fig:intro}a) plots $\chi$ normalized by its value $\chi_0 \equiv \chi(V=0)$. As shown in Fig.~(\ref{fig:intro}a), $\chi$ reduces quickly with increase of $V$ and becomes vanishingly small for $V \ge V_1$, here $V_1 \sim 0.8$. This destruction is very similar to the decay of the commensurate CDW order in 1T-\chem{TiSe_2} with \chem{Cu}-intercalation~\cite{Cava0}. 

In order to analyze the CDW order within domains which are out of phase with each other we define the CDW amplitude as $ \vert \chi \vert \equiv \langle \vert \chi_i \vert \rangle=N^{-1} \langle \sum_i \vert \chi_i \vert\rangle_c$. The non-vanishing nature of the CDW amplitude $\vert \chi \vert$ even for $V > V_1$ indicates the existence of short range CDW order within each domain as evident in Fig.~(\ref{fig:intro}a), which survives up to rather large values of $V$. The superconducting correlation, on the other hand, is signaled by the off-diagonal long range order, defined as:
\begin{equation}
\Delta_{\rm OP} = U \sqrt{\langle c^\dagger_{i \uparrow} c^\dagger_{i \downarrow} c_{j\downarrow} c_{j \uparrow} \rangle}
\end{equation} 
as $\vert i-j \vert \rightarrow \infty.$ We find that the $\Delta_{\rm OP}$ starts increasing rapidly at $V \sim V_1$, as seen in Fig.~(\ref{fig:intro}a), where the global CDW order gets destroyed.

We also consider another limit of mean-field decomposition, where the CDW amplitude $\chi_i$ in Eq.~(\ref{densityform}) is forced to zero for all sites $i$.
\begin{align}
\mathcal{H}_{\rm sSC}=&\sum_{\langle i,j \rangle, \sigma} t_{ij} \left(c^{\dagger}_{i  \sigma} c_{j \sigma} + h.c.\right) + \sum_{i, \sigma} (V_i-\mu-\frac{|U|}{2}\rho_0(i))n_{i \sigma} \nonumber \\ &+\frac{W}{2}\sum_{\langle i,j \rangle,\sigma} \rho_0(j) n_{i, \sigma} +\sum_{i} (\Delta_i c_{i\uparrow}^\dagger c_{i \downarrow}^\dagger +h.c.)
\label{sSCmodel}
\end{align}
Above Hamiltonian helps a justified comparison with results from $\mathcal{H}^{\rm CDW}_{\rm sSC}$. We emphasize that the determination of the spatial profile of $\rho_0(i)$ does not involve self-consistency in $\langle n_{i} \rangle$. On the other hand, a self-consistent determination of $\langle n_{i} \rangle$ invariably gives rise to the charge order in Eq.~(\ref{densityform}) and is excluded in the analysis of $\mathcal{H}_{\rm sSC}$. In fact, the spatial profile of local density $\rho_0(i)$ obtained from the solution of the Anderson model ($U=0,W=0$ in the Hamiltonian in Eq.~(\ref{EHM}))

is used as its input for an accelerated self-consistency in pairing amplitude using Hamiltonian Eq.~(\ref{sSCmodel}). We used the Bogoluibov-de Gennes (BdG) transformation to diagonalize $\mathcal{H}_{\rm sSC}$ or $\mathcal{H}^{\rm CDW}_{\rm sSC}$ separately at $T=0$, following Ref.~\onlinecite{PdGbook} and self-consistently solve the BdG equations. We used different methods for accelerating the convergence in self-consistency, including linear mixing, Broyden method and modified Broyden method~\cite{ModBroyd}, and their combinations.

We now turn to discuss the key results from the clean system first. For this purpose we solve the underlying `gap-equations' and the density equation in the momentum space in a manner similar to the BCS theory, upon including the competing orders.

\section{Phase diagram for clean system}

\begin{figure}
      \includegraphics[width=8.5cm,height=10.0cm,keepaspectratio]{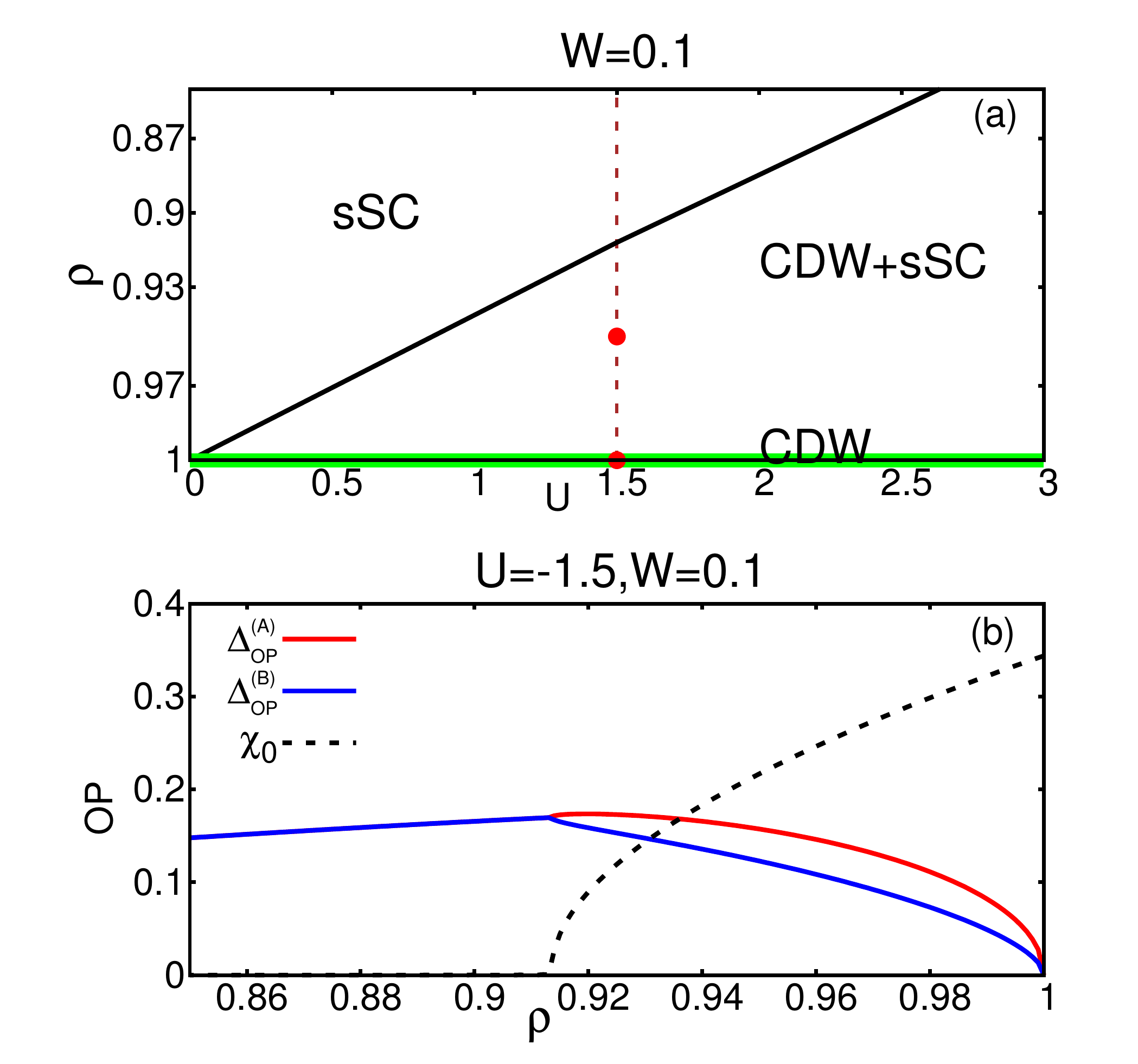}
      \caption{(Color Online) Panel \textbf{(a)} shows the $T=0$ phase diagram of the clean system described by the Hamiltonian in Eq.~(\ref{Comp}) in the $(U,\rho)$-plane, for a fixed $W=0.1$. The green horizontal line along x-axis indicates CDW ground state at half-filling ($\rho=1$). Away from the half-filling, a mixed phase containing both the CDW and sSC orders is stabilized as ground state up to a critical density $\rho_c\equiv \rho_c(U,W)$. Beyond $\rho_c$, only a sSC ground state arises. The red dots indicate the parameter values for which the effect of disorder on the interplay of the two orders are addressed in this study. Panel \textbf{(b)} shows the evolution of the charge modulation amplitude and the pairing amplitude with density in the clean system for a fixed set of interaction parameters: $U=-1.5$ and $W=0.1$. The CDW order parameter (shown in black) reduces continuously from $\rho=1$ and finally vanishes at $\rho_{c} \sim 0.91$. Pairing amplitude on sublattice A, $\Delta^{\rm(A)}_{\rm OP}$ (shown in red), and the same on sublattice B, $\Delta^{\rm (B)}_{\rm OP}$ (shown in blue) disappear at $\rho=1$. Doping the system slightly away from the half-filling modulates pairing amplitude in space. Such modulation goes away and only uniform superconducting pairing amplitude results for $\rho \leq \rho_c$.}  
      \label{fig:phase_dia} 
   \end{figure}

A clean system respects the lattice translation symmetry. Since the checkerboard spatial modulation of CDW breaks this symmetry into two sub-lattices,
we transform the Hamiltonian in A- and B-sublattices and subsequently work in Fourier space to analyze results for each of the sub-lattices. We include the details of the formalism in Appendix A.

The resulting phase diagram for the clean system is shown in Fig.~(\ref{fig:phase_dia}a). A CDW phase without any sSC order emerges as the ground state at half-filling due to non zero $W$, as mentioned already. At $\rho \neq 1$ the ground state shows a coexisting CDW and sSC order. The CDW and sSC live together up to a threshold values of density $\rho_c$ (where $\rho_c=\rho_c(U,W)$), beyond which the sSC order wins over the density modulations.
\begin{figure}
      \includegraphics[width=8.5cm,height=10.0cm,keepaspectratio]{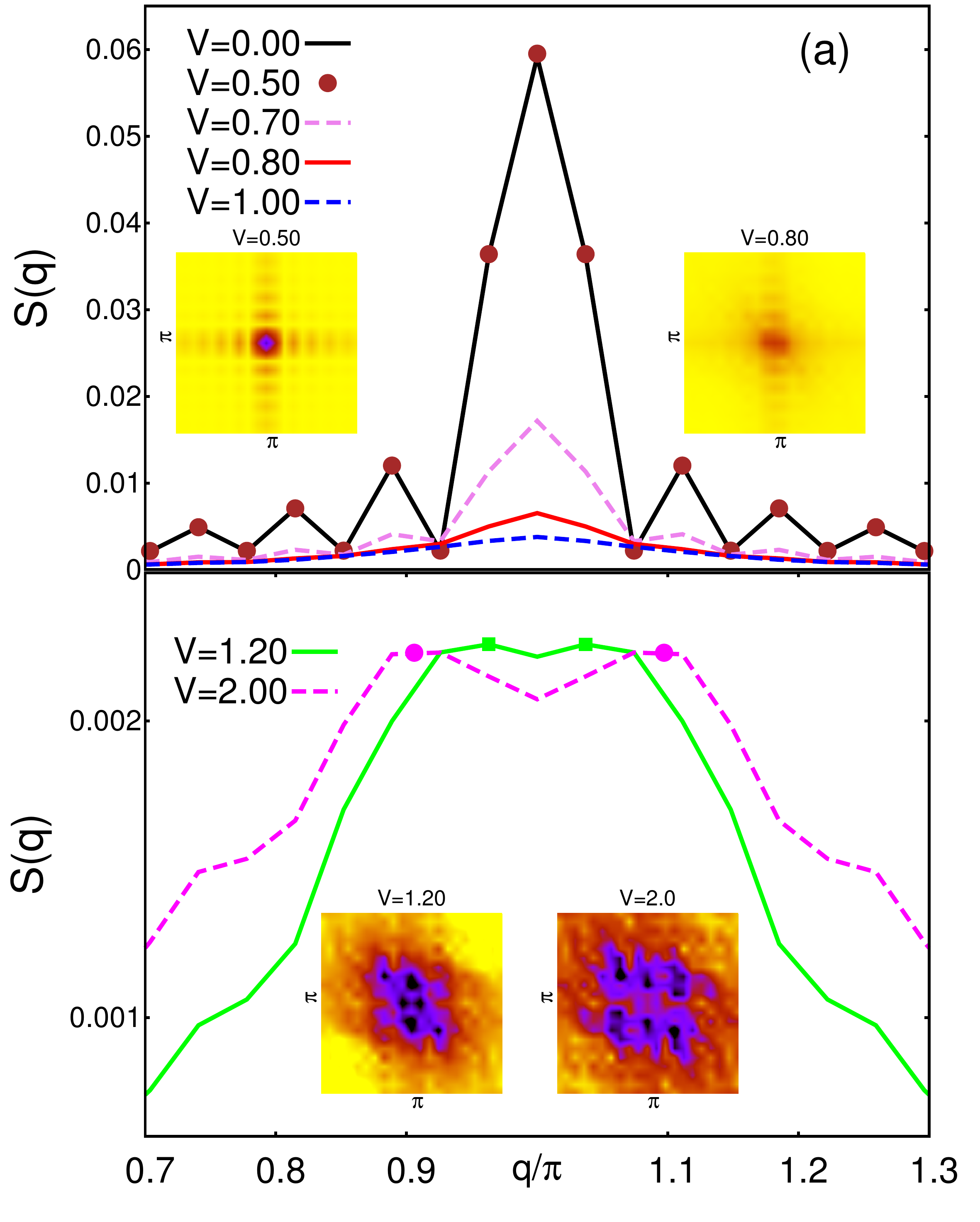}
      \caption{(Color Online) Panel {\bf (a)} shows the evolution of $S(\bf{q})$ for different $V$. The trace at $V=0$ features a sharp peak at $\bf{q}=(\pi,\pi)$, which falls off rapidly by $V_1 \sim 0.8$. Increasing $V$ from $V_1\sim 0.8$ to $V_2\sim 2$ broadens the peak of $S(\bf{q})$ at $\bf{q}=(\pi,\pi)$, indicating presence of short ranged CDW puddles. The 2D color-density plots, shown as insets, feature only single peak at $\bf{q}=(\pi,\pi)$ for $V=0.5$ and $V=0.8$. Panel {\bf (b)} shows a weak splitting and spreading of the structure factor peak for $V \gtrsim V_1$ (shown here for $V=1.2$ and $V=2.0$). The 2D color-density plots are shown as inset, indicates large broadening and splitting of the structure factor peak, at these $V$'s.}  
      \label{fig:struct_fact}
   \end{figure}
The variation of the order parameters with $\rho$ for a fixed $U=-1.5$ and $W=0.1$ (in the unit of t=1.0) is shown in Fig.~(\ref{fig:phase_dia}b). The CDW order $\chi_0$ (shown in black) falls of smoothly as density decreases away from half-filling and vanishes around $\rho_c\sim0.91$ for our chosen parameters. The sSC pairing of the two sublattices is suppressed due to non-zero $W$. The sSC pairing at A-sublattice $\Delta^{\rm (A)}_{\rm OP}$ and B-sublattice $\Delta^{\rm (B)}_{\rm OP}$ increases with decrease in density, because CDW order weakens away from half-filling. Interestingly, the sSC pairing amplitude in two sublattices are different, signifying that the CDW component induces a modulating sSC pairing amplitude in space. This modulation in sSC pairing amplitude is lost beyond $\rho_c$ because the CDW order vanishes.

The half-filled attractive Hubbard model ($W=0$), which also has the coexisting sSC and CDW orders, does not yield such a modulating sSC pairing amplitude. This is because, at half-filling this model has a particle-hole symmetry, and a loss of this away from half-filling gives rise to such modulation.

The effect of disorder on a sSC phase had been explored in the past~\cite{GTR},\footnote{We have checked that the additional nearest neighbor repulsion doesn't change the qualitative physics}. We focus below on the parameter values shown as red dots in Fig.~(\ref{fig:phase_dia}a) and study the disorder dependence of the coexisting CDW and sSC phases. Most of the results presented in the following section are for $\rho=1$ though some subtleties away from half-filling are discussed in Appendix B.

\section{Interplay between SC and CDW in the presence of disorder}
In order to analyze the interplay in disordered systems we report below our results from the two models ${\cal H}^{\rm CDW}_{\rm sSC}$ and ${\cal H}_{\rm sSC}$ for parameters shown in Fig.~(\ref{fig:phase_dia}), namely, $U=-1.5,W=0.1$ (in the unit of $t=1$) and for $\rho =1.0$ and $0.95$ respectively. We used system sizes up to $N=54 \times 54$. We average all observables over $15$-$20$ independent realizations of disorder for each strength of $V$. We found that the ground state energy of the model $\mathcal{H}^{\rm CDW}_{\rm sSC}$ remains below that of $\mathcal{H}_{\rm sSC}$ for all $V$. Moreover, the ground state energies of these two Hamiltonians approach the same value as $V$ increases, implying that CDW loses significance for strong disorder~\footnote{We have verified that the length scale of CDW fluctuations at large $V$($>V_2$) becomes comparable to lattice spacing}. Similar results for the free energy were obtained for the hole doped ($\rho = 0.95$) system studied.

\begin{figure}
      \includegraphics[width=8.5cm,height=10.0cm,keepaspectratio]{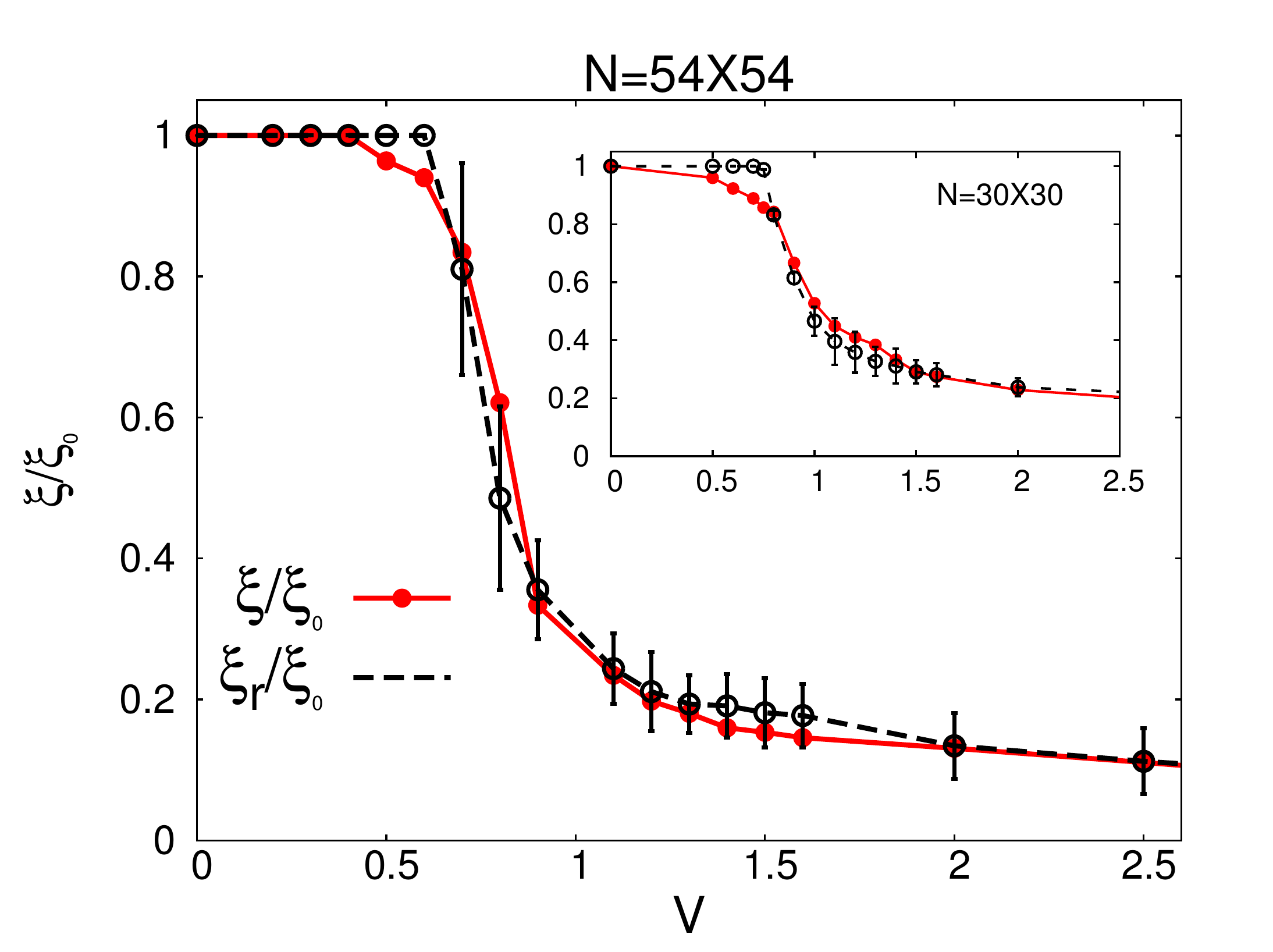}
      \caption{(Color Online) Comparison between the length scale of CDW correlations, calculated from two independent methods: (A) The red curve presents its evolution with $V$ as calculated from the width of the peak of $S(\bf{q})$ at $\bf{q}=(\pi,\pi)$ (denoted as $\xi$). (B) The dashed line depicts the evolution of the typical cluster size (denoted as $\xi_r$) of charge modulation as a function of $V$ using statistical analysis (described in Appendix C). Here, $\xi_{0}\equiv \xi_{V=0}$, and the calculations are done on a $54 \times 54$ system. Both these traces show nearly identical trend. The Inset shows the same length-scales from the two methods, but for a smaller system, a $N = 30 \times 30$ lattice. The apparent saturation of both $\xi$ and $\xi_r$ occurs at a slightly higher value of $V$ on the smaller system, which is more severely affected by finite size effects.}  
      \label{fig:xi}
   \end{figure}

\subsection{Structure Factor}
Charge modulation due to CDW order is characterized by the structure factor, which is routinely measured in x-ray diffraction experiments~\cite{DWinSolids}. In order to define this, we first calculate the charge correlation function $\mathbb{C}(\mathbf{r})$, given by:
\begin{equation}
\mathbb{C}( \mathbf{r_i}-\mathbf{r_j} )= \left\langle \sum_{\sigma,\sigma^\prime} \left( \hat{n}_{i \sigma} -\rho \right)\left( \hat{n}_{j \sigma^\prime} -\rho \right) \right\rangle,
\end{equation}
here the average is taken over disorder realizations, as well as over all possible combination of i and j which leave $\vert r_i-r_j \vert$ unaltered. 
The structure factor is defined as the Fourier transform of $\mathbb{C}(\mathbf{r})$ as:
\begin{equation}
S(\mathbf{q})=\frac{1}{N}\sum_{r} e^{i \mathbf{q}.\mathbf{r}} \mathbb{C}(r).
\end{equation}

The dominant peak in the structure factor in momentum-space occurs at the ordering wave-vector of the density modulation. Our result of $S(\mathbf{q})$, shown in Fig.~(\ref{fig:struct_fact}), demonstrates that the peak position of $S(\mathbf{q})$ does not change with $V$, up to $V=1.0$ (which is past $V_1$, see Fig.~(\ref{fig:intro})), though  the peak gets broadened and the intensity at $(\pi,\pi)$ decreases with the increase in disorder. In fact, $S(\pi,\pi)$ follows the same trend with V as $\chi$ does, essentially vanishing (modulo the finite-size effects) at $V_1$ indicating the loss of commensurate CDW order.

What footprints of incommensuration do we see in our simulations?
For $V=1.2$ and $V=2.0$, we see that the single peak at $(\pi,\pi)$ gets suppressed and multiple weak peaks appear at $\bf q$ points slightly away from $(\pi,\pi)$ as shown in the main panel of Fig.~(\ref{fig:struct_fact}b) and also in the right inset of the same panel. This is in broad qualitative agreement with the STM study on 1T-\chem{TiSe_2}~\cite{YanSTM}.

As mentioned, peak(s) in $S(\bf{q})$ smears out and widens with $V$, implying that the correlation length associated with the checkerboard CDW order reduces with $V$~\cite{DWinSolids}. 
The full width at half-maximum (termed $\gamma$ here) of the peak of $S(\bf{q})$ at $\bf{q}=(\pi,\pi)$ naturally defines such correlation length as: $\xi \sim 1/\gamma$. We have presented the $V$-dependence of this $\xi$ in the Fig.~(\ref{fig:intro}a) which falls off with $V$, but at a slower rate than $\chi$ (See Fig.~(\ref{fig:intro}a)), with an apparent saturation around $V_2 \approx 2$. Note that the smallest value that $\xi$ can attain is of the order of lattice spacing and our saturation value at large $V$ is consistent with that. We have also performed an independent statistical analysis of the typical CDW `puddle' sizes for all different $V$, by generating statistics from all independent disorder realizations. The $V$-dependence of this puddle-size, which we denote as $\xi_r$, follows closely the behavior of $\xi$ within numerical error as shown in Fig.~(\ref{fig:xi}), yielding confidence in our extraction of this length scale.

Thus our calculations confirm that the CDW fluctuations persist beyond $V_1$, even though the global ordering is wiped out. The x-ray diffraction experiments are routinely carried out on CDW bearing materials. 
For example, on \chem{Cu_xTiSe_2}~\cite{Abbamonte} such studies reveals that with increasing $x$ $S(\mathbf{q})$-peak broadens and hence the CDW correlation length decreases. Similarly, 2H-\chem{NbSe_2} shows depletion of peak as well as broadening compared to the pristine sample with \chem{Mn} or \chem{Co} intercalation~\cite{Arpes2HNbSe2}. Similar observations are also made for high-$T_c$ material \chem{HgBa_2CuO_{4+x}} with oxygen doping~\cite{DisorderCDWHTSc}.

\begin{figure}
      \includegraphics[width=12.0cm,height=16.0cm,keepaspectratio]{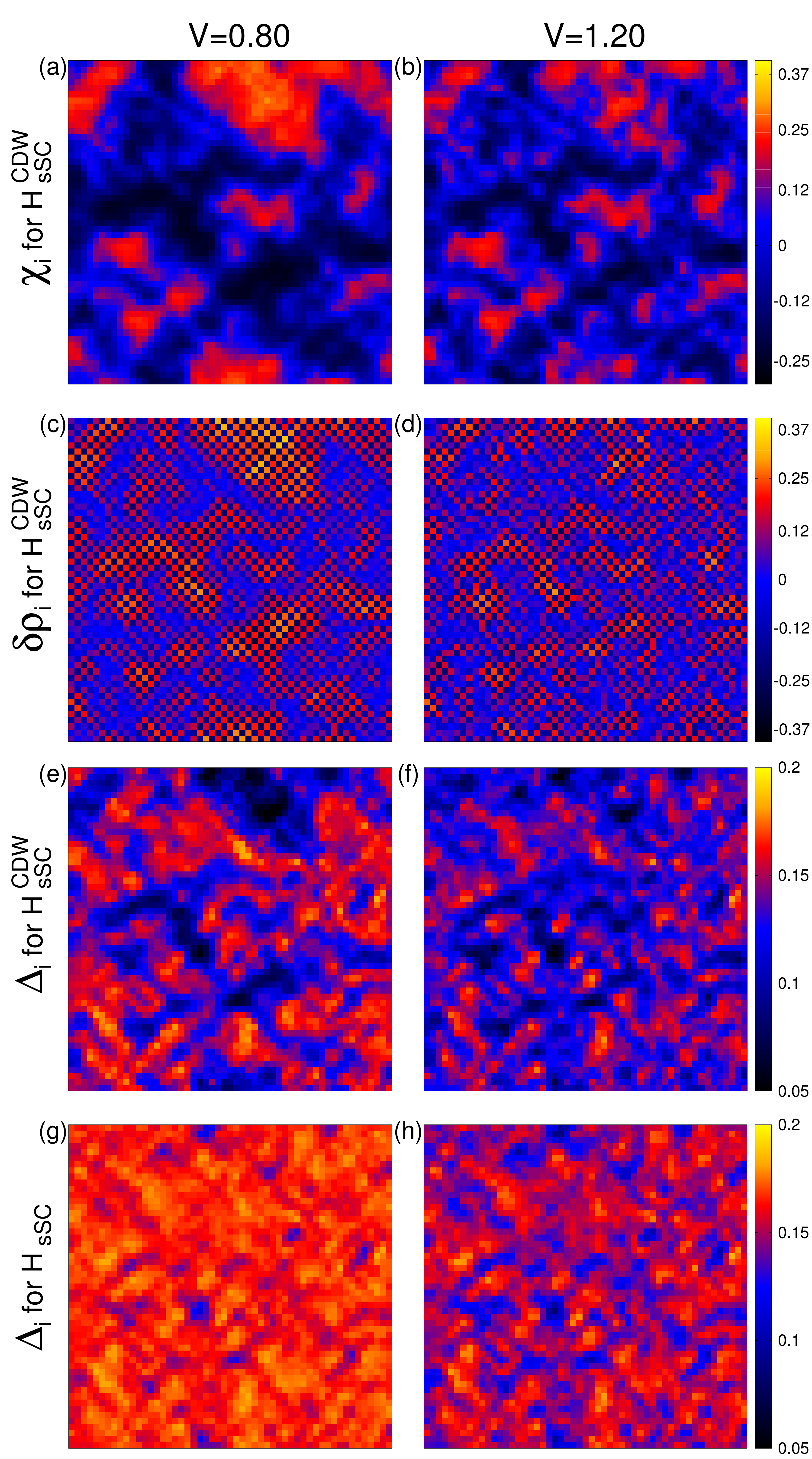}
      \caption{(Color Online) Panel \textbf{(a)} shows the spatial profile of CDW amplitude $(\chi_i)$ for $V=0.8$, representing relatively large regions of positive (yellow) and negative (blue/black) islands. Panel \textbf{(b)} is representation of silmilar data as in panel \textbf{(a)}, but for $V=1.2$, which exhibits more fragmented structure of $\chi_i$. Panel \textbf{(c)} depicts spatial profile of the density modulation for $V=0.8$ which is large in regions of both positive and negative $\chi_i$. Panel \textbf{(d)} is similar to panel \textbf{(c)}, but for $V=1.2$. Panel \textbf{(e)} shows spatial profile of the sSC pairing amplitude ($\Delta_i$) arising in $\mathcal{H}^{\rm CDW}_{\rm sSC}$ model at $V=0.8$, which demonstrates its spatial separation from the regions of strong density modulation when compared with panel \textbf{(a)}. The panel \textbf{(f)} shows the same quantity as in panel \textbf{(e)} but for $V=1.2$. In panel \textbf{(g)}, we present $\Delta_i$ for $\mathcal{H}_{\rm sSC}$ at $V=0.8$ where we find a much weaker variation in it than from the model $\mathcal{H}^{\rm CDW}_{\rm sSC}$, shown in panel \textbf{(e)}. Finally, panel \textbf{(h)} shows the same results as in panel \textbf{(g)}, but for $V=1.2$.}  
      \label{fig:LOP}
\end{figure}

\subsection{Spatial distribution of local orders}
While the global nature of the interplay between the two competing orders is already depicted in Fig.~(\ref{fig:intro}), a deeper insight on the spatial reorganization of order parameters is best obtained by studying the disorder dependence of these profiles. For this purpose, we first investigate on a specific realization of disorder the local orders arising from ${\cal H}^{\rm CDW}_{\rm sSC}$ at $\rho=1$ and $V=0.8$ ($\approx V_1$). We see in Fig.~(\ref{fig:LOP}a) that the staggering amplitude $\chi_i$, which is homogeneous in the clean limit, gets segmented into regions of its positive (yellow) and negative (black) valued `puddles'.

The typical size of these puddles is given by $\xi$, whose $V$-dependence is already illustrated in Fig.~(\ref{fig:xi}). Such fragmentation was found to occur by occasional shifting of the periodicity of the CDW modulation, and the corresponding $\pi$-phase shifted density modulation is shown in Fig.~(\ref{fig:LOP}c). 

Had these phase slips been the only mechanism for the destruction of the global CDW ordering, it is better called ``discommensuration"~\cite{McMillan0,McMillan1,Littlewood0,ctoICDW}, rather than an incommensuration. A lock-in phase transition from commensurate to incommensurate CDW state was studied in the past within Ginzberg-Landau type field theoretic models, leading to a phenomenon termed `discommensuration'. A true incommensuration results in a splitting (or a small shift) of ordering wave-vector as found in Ref.~[\onlinecite{YanSTM}]. Interestingly, in experiment such splitting of $S(\mathbf{q})$-peak is also accompanied by formation of phase shifted domains. Our simulation shows unambiguous identification of phase-shifted domains, whereas, we see weak signatures of the splitting of the structure factor peak in our numerical results, as discussed in the previous section. Further, recent explorations of charge modulation in high temperature cuprate superconductor: BSCCO, proposes novel mechanism through which discommensuration can give rise to incommensuration~\cite{Mesaros}. Having noted these, we loosely use `discommmensuration' and `incommensuration' more or less synonymously in the rest of our manuscript.

Notice that the density modulation is equally strong on both the positive and negative puddles of local CDW ordering. The local sSC pairing amplitude $\Delta_i$ is shown in Fig.~(\ref{fig:LOP}e) which is weak on strong CDW puddle -- both the positive and negative ones, indicating a spatial anticorrelation of the presence of superconductivity and that of charge modulations. In contrast, Fig.~(\ref{fig:LOP}g) presents the self-consistent $\Delta_i$ obtained from the ${\cal H}_{\rm sSC}$ for the same disorder, illustrates only minor depressions in sparse locations of the system. Thus, the superconductivity remains strong and nearly homogeneous in the absence of competing charge order. 

Similar results for $V=1.2$ in Fig.~(\ref{fig:LOP}b,\ref{fig:LOP}d) shows that CDW islands disintegrate into smaller ones with the increase of $V$, whereas, the sSC pairing amplitude begins to become dominant, as seen from Fig.~(\ref{fig:LOP}f). However, the spatial anticorrelation between the two orders persists. The spatial profile of $\Delta_i$ from ${\cal H}_{\rm sSC}$ in Fig.~(\ref{fig:LOP}h) continues to differ from that in the `coexistence' model. This confirms that the presence of local CDW at $V=1.2$ alters the profile of $\Delta_i$, though the global CDW has already collapsed. We have verified that upon increasing $V$ the CDW islands shrink further and by $V_2 \sim 2$ they become of the size $\sim 2$ to $3$ lattice spacing. In addition, the spatial profile of $\Delta_i$ calculated from ${\cal H}^{\rm CDW}_{\rm sSC}$ and ${\cal H}_{\rm sSC}$ tend to become identical at these disorders.

\begin{figure}
      \includegraphics[width=9.0cm,height=16.0cm,keepaspectratio]{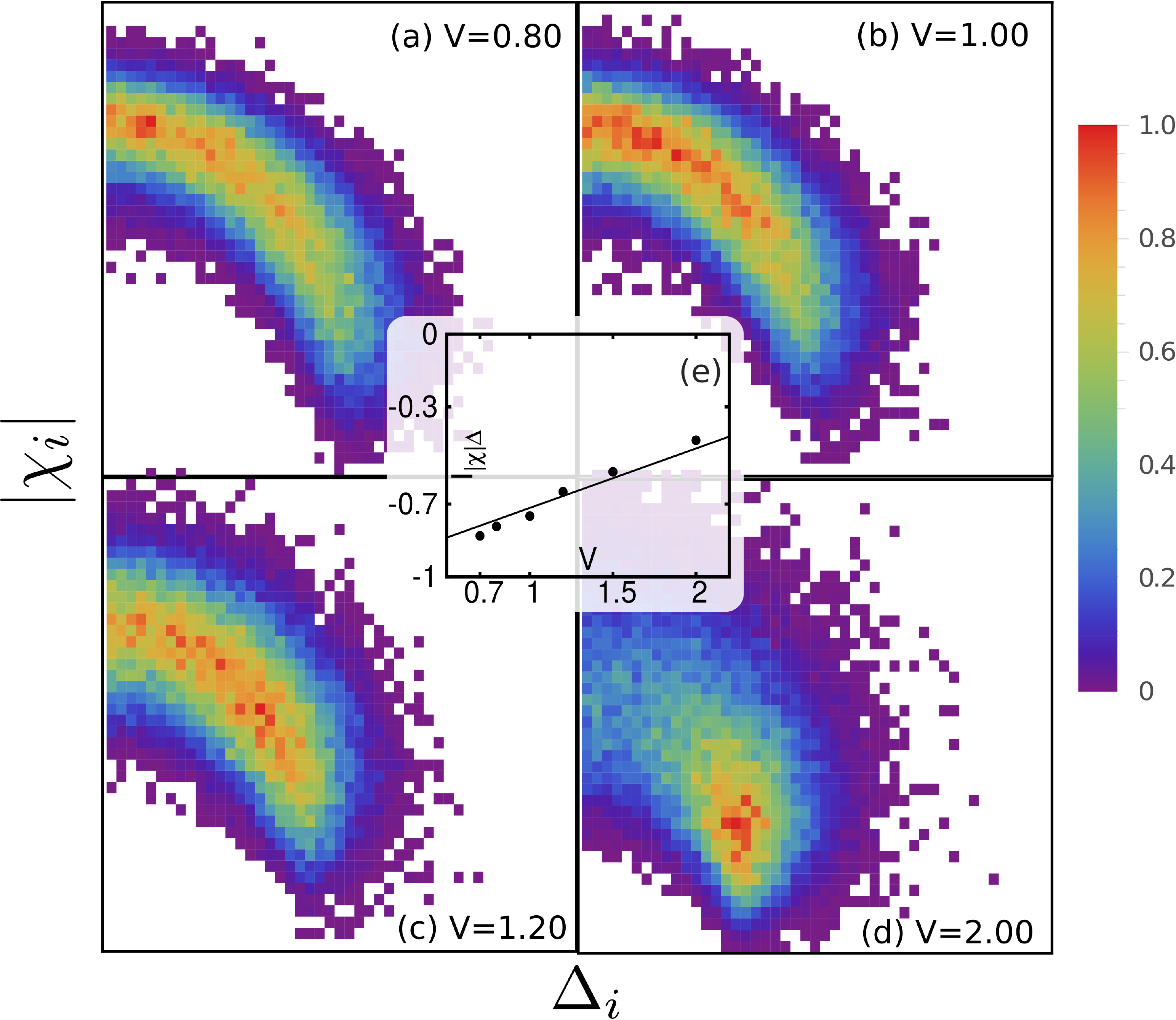}
      \caption{(Color Online) Panels (\textbf{a})-(\textbf{d}) presents the cross-correlation histogram between $\Delta_i$ and $\vert \chi_i \vert$ for $V=0.8, 1.0, 1.2 \text{ and } 2.0$ respectively, showing the inverse spatial correlations between the two quantities. Panel (\textbf{e}) shows that the evolution of $I_{\vert \chi \vert \Delta}$ as a function of $V$ is largely linear.}
      \label{fig:anticorr}
\end{figure}
We highlight here that the spatial anticorrelation described above is qualitatively similar to the experimental findings of the \chem{Cu}-intercalated 1T-\chem{TiSe_2}~\cite{YanSTM}. To emphasize this observation, we present in Fig.~(\ref{fig:anticorr}) the scatter plots of $\vert \chi_i \vert$ and $\Delta_i$ for various strengths of disorder. The anticorrelation is evident from the negative slope of the scatter plots in each panel, though the increase in $V$ broadens the scattering of data signaling a weakening of the anticorrelation. We further quantify this by estimating the cross-correlator, $I_{\vert \chi \vert \Delta}$, as follows:
\begin{equation}
I_{\vert \chi \vert \Delta}=\frac{1}{N} \sum_{i=1}^N\frac{\left(  \vert \chi_i \vert - \langle \vert \chi \vert\ \rangle\right)\left( \Delta_i - \langle \Delta \rangle \right)}{\sigma_{\vert \chi \vert} \sigma_{\Delta}}
\end{equation}
where $\sigma_{\vert \chi \vert}$ is the standard deviation of $\vert \chi_i \vert$ and $\sigma_{\Delta}$ is the same for the sSC pairing amplitude. $\langle \hdots \rangle$ denotes averaging over all the site, as well as over independent disorder configurations. A perfect anticorrelation yields $I_{\vert \chi \vert \Delta}=-1$. We have shown the evolution $I_{\vert \chi \vert \Delta}$ with $V$ in Fig.~(\ref{fig:anticorr}e).
There is strong spatial anti-correlation between the two orders around $V\approx0.8$, which gradually weakens as V is increased.

What's the mechanism behind the the transition from commensurate to incommensurate CDW, which coincides with nucleating superconductivity in the domain walls of CDW puddles? Two terms of the Hamiltonian ${\cal H}$ of Eq.~(\ref{EHM}) contribute to charge inhomogeneities -- (a) the interactions that lead to (long-range) modulated charge density, and (b) the uncorrelated disorder, which causes the local density to respond to its spatial profile. For  $V \leq V_1$, disorder energy remains weak and can at best alter the charge modulation pattern of the ground state, over a correlation length $\xi$ -- the typical size of the puddles. The number of puddles with opposite polarity remain statistically similar, causing $\chi$ to vanish at $V_1$. As $V$ increases, $\xi$ decreases and local density starts responding only to local $V_i$, as the relevance of interactions goes down compared to disorder. As a result, the self-consistent spatial density approaches to that of the underlying Anderson's model~\cite{PWA} of disorder (in which $U = 0 = W$) by $V \sim V_2$. Any remaining ultra short-range CDW is irrelevant beyond this point.

On the other hand, the survival of superconductivity is not very sensitive to weak fluctuations in local density and inhomogeneous sSC order thus emerges in the `domain walls' between CDW islands. Our study of density of states (DOS) sheds more light on why sSC pairing develops in the domain region, and will be discussed in Sec.~(\ref{sec:DOS}). Superconductivity cannot nucleate within CDW puddles by the very construction of the model, whose parameters ensure that sSC is only a sub-dominant order in such regions of low disorder.

We proceed below to investigate the effect of above local reorganization on the observables.

\subsection{Distribution of local observables and density}
\begin{figure}
      \includegraphics[width=8.5cm,height=10.0cm,keepaspectratio]{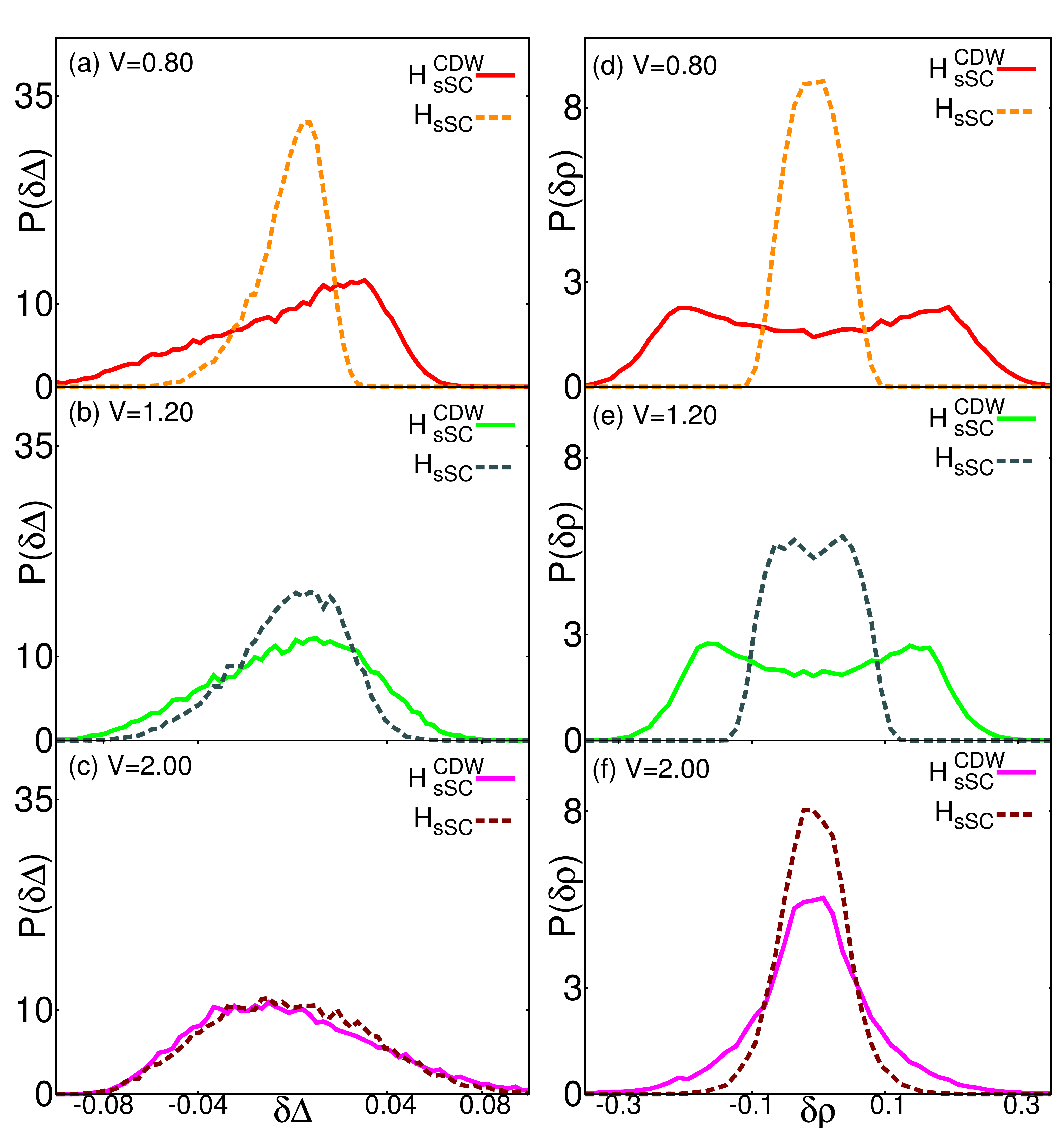}
      \caption{(Color online) The left panels present the distribution of the fluctuations in the sSC pairing amplitude, defined as: $\delta \Delta_{i}=(\Delta_{i}-\langle \Delta \rangle)$ for models $\mathcal{H}^{\rm CDW}_{\rm sSC}$ and $\mathcal{H}_{\rm sSC}$ for (\textbf{a}) $V=0.80$ (\textbf{b}) $V=1.20$, and (\textbf{c}) $V=2.00$. A strong difference between $P(\delta \Delta)$ from the two models persists in the disorder window $V_1$ to $V_2$. The two models show similar features beyond $V_2$. The right panels shows the corresponding plots for the distribution of the density modulation. The bi-modal nature of the distribution from the model $\mathcal{H}^{\rm CDW}_{\rm sSC}$, as well as the difference between the results from the two models vanish smoothly around $V_2 \sim 2.0$.}  
      \label{fig:dist_OP}
   \end{figure}

We next confine our attention to the fluctuation in the self-consistent pairing amplitude from the two models $\mathcal{H}^{\rm CDW}_{\rm sSC}$ and $\mathcal{H}_{\rm sSC}$. We define this fluctuation $\delta \Delta_{i}$ by subtracting the mean value: $\delta \Delta_{i}=(\Delta_{i}-\langle \Delta \rangle)$. Removal of this baseline of average pairing amplitude is necessary for a fair comparison between the two models, because, $\langle \Delta \rangle$ differs in the two calculations due to the inclusion and exclusion of self-consistency in $\rho_i$ in  $\mathcal{H}^{\rm CDW}_{\rm sSC}$ and $\mathcal{H}_{\rm sSC}$. We show the distribution of $\delta \Delta_{i}$, i.e. $P(\delta \Delta)$, on the left panels of Fig.~(\ref{fig:dist_OP}). If the local CDW had no effects on the disordered sSC, $P(\delta \Delta)$ from the two models should be identical. Our result suggests that they remain significantly different for $V \gtrsim V_1$. The difference shrinks smoothly by $V \sim V_2$, implying that short- range CDW fluctuations become irrelevant. The window of disorder, $1.5 \le V_2 \le 2$, that marks the disappearance of the differences between the two model, is consistent with the value at which the correlation length $\xi$ achieves its saturation, as seen in Fig.~(\ref{fig:intro}a).

Not only the presence of local CDW order makes the $P(\delta \Delta)$ wider compared to when CDW is absent, it also features a negative skewness arising due to the underlying interplay. This is easily comprehended: We found in the previous section that the CDW resides in regions where sSC pairing amplitude is weaker. As a result, the sites where $\Delta_i$ is small in model $\mathcal{H}_{\rm sSC}$, becomes even smaller by accommodating the charge modulations locally, whereas the regions of large $\Delta_i$ remain unaffected since local CDW amplitude is negligibly small on those sites. The distribution for the model $\mathcal{H}^{\rm CDW}_{\rm sSC}$ becomes progressively symmetric as $V \rightarrow V_2$, signifying that the fluctuation of sSC pairing is due to the presence of random disorder and not from any competing order. 

The distribution of the density modulation, $\delta \rho(i)=\rho(i)-\rho_0(i)$, is also shown in Fig.~(\ref{fig:dist_OP}d-\ref{fig:dist_OP}f). At $V=0.8$ the bi-modality of $P(\delta \rho)$ obtained from $\mathcal{H}^{\rm CDW}_{\rm sSC}$ is evident, whereas, it is absent in results from $\mathcal{H}_{\rm sSC}$. The bi-modality is a clear signature of the surviving CDW fluctuations~\footnote{We have removed the Anderson Model density from the local density to minimize the bi-modality arising from the disorder.}. With increasing $V$, the difference between the solid and the dashed traces reduces smoothly and disappears beyond $V_2$.

\subsection{Density of states} \label{sec:DOS}
\begin{figure}
      \includegraphics[width=9.0cm,height=10.0cm,keepaspectratio]{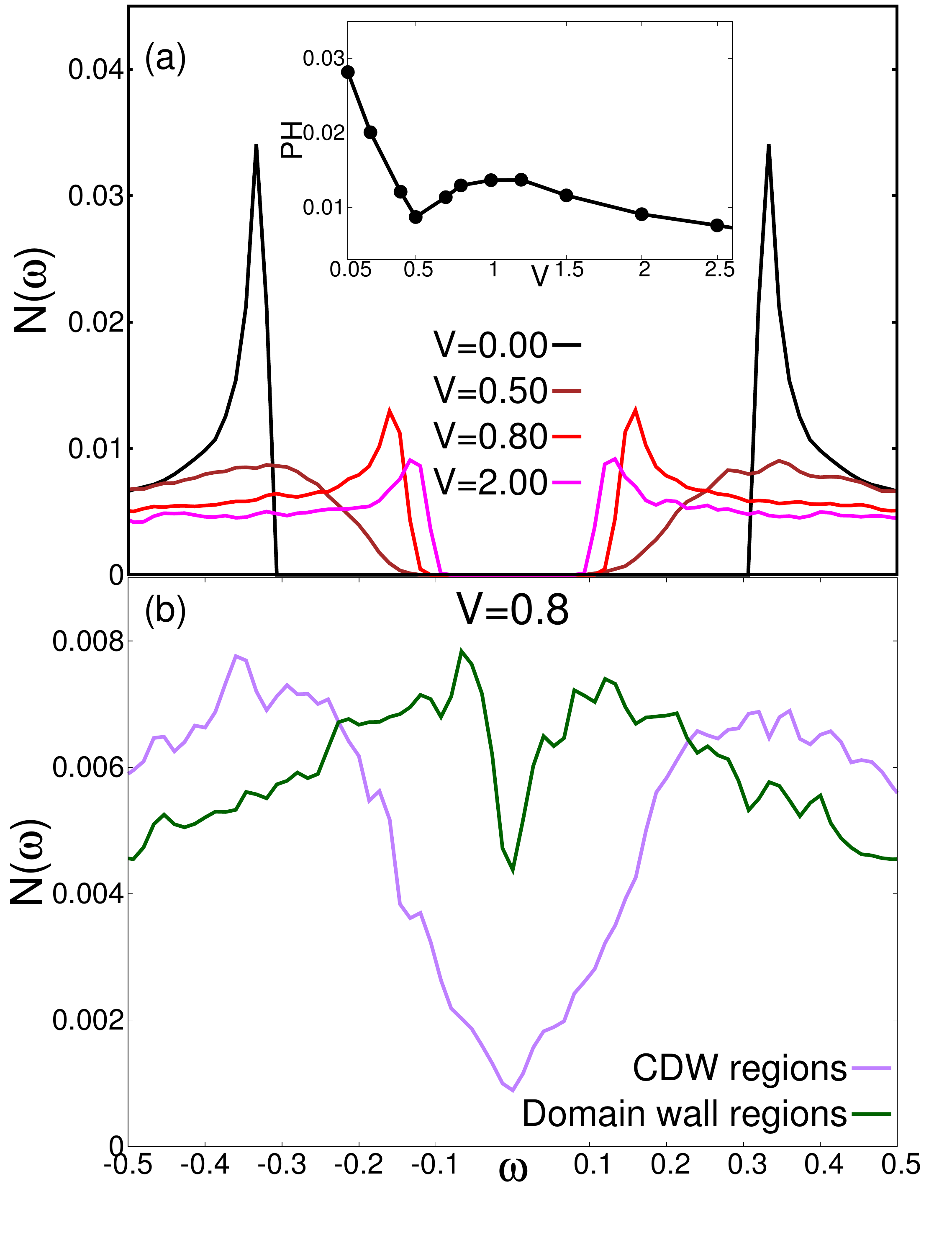}
      \caption{(Color online) The density of states (DOS) averaged over all sites and also over disorder configurations for several values of $V$ are shown in the panel (\textbf{a}). These results show an initial quick fall of the peak in the DOS at the gap edge, which is followed by the development of superconducting coherence peaks at larger disorder strengths. Inset of panel (\textbf{a}) shows the variation of peak height (PH) near the gap edge with $V$. Panel (\textbf{b}) presents the coarse grained local density of states on strong CDW-puddles ($\vert \chi_i \vert >0.2$) and on the domain between adjacent such puddles where, $\vert \chi_i \vert <0.02$. The large weight of low energy DOS in those domain walls makes it easier for superconductivity to emerge on those domain walls.}  
 \label{fig:TDOS}
   \end{figure}
The interplay of competing orders in ground state has consequences for density of states (DOS) $N(\omega)$, as both superconducting and charge orders leave distinctive footprints on it. We present $N(\omega)$ averaged over all sites in a half-filled system in Fig.~(\ref{fig:TDOS}a) for different strengths of disorder. 
 
The strong peak near the gap edge in $N(\omega)$ at $V=0$ in Fig.~(\ref{fig:TDOS}a), arises from the splitting of the van-Hove singularity, due to the emergence of charge modulations. The magnitude of this gap, arising from the CDW order, is consistent with $\chi_0$ and is given by $(U+Wz)\chi_0/2$, where the coordination number, $z=4$ for the square lattice.

Note that the subdominant superconducting order has no role in $N(\omega)$ for the clean system ($V=0$). As $V$ increases, charge modulation weakens, causing a rather sharp fall of the height of the CDW peak at the gap edge, and states below the gap edge begin to populate. By $V=0.5$, the $N(\omega)$ features only a broad hump near the gap edge (of $V=0$ result) as seen in Fig.~(\ref{fig:TDOS}a), which is the remnant of the depleted split-peak of Van-hove singularity. Interestingly, for $V \gtrsim 0.8(\sim V_1)$, even though the global CDW is destroyed, $N(\omega)$ develops a new peak.

This peak (though weak in the presence of disorder) is due to the emerging superconductivity in the domain regions separating strong CDW puddles, as discussed before. The magnitude of this gap is consistent with $\Delta_0$, and indeed has origin different from the charge modulation. They are are due to (weak) superconducting coherence, in the presence of disorder. Note that, the s-wave superconductivity is not sensitive to moderate $V$, due to Anderson's theorem~\cite{AndTheo}. The resulting $N(\omega)$ evolves at larger $V$s in a manner, which is consistent with what is reported in the literature~\cite{GTR}. 
 
The above discussion highlights an intriguing non-monotonic evolution of peak height (PH) at the gap edge in DOS with $V$ which is explicitly shown in the inset of Fig.~(\ref{fig:TDOS}a). The first dip around $V_1$ is easily discernible, and can be addressed in scanning tunneling spectroscopy experiments. PH shows a non-monotonic behavior beyond the initial dip as well. Its increase beyond $V_1$ occurs in a disorder range that is rife with fluctuating charge modulation coexisting with superconductivity. The final fall of PH for large disorder $V \sim V_2$ is due to the loss of superconducting coherence in a disordered superconductor~\cite{GTR}.  

In order to probe the nucleation of superconductivity within `domain walls' we perform a simple calculation as follows. We force $\Delta_i = 0$ at all sites $i$ in the Hamiltonian $\mathcal{H}^{\rm CDW}_{\rm sSC}$ and calculate the local density of states (LDOS) averaged in two different (coarse-grained) regions: (a) On sites belonging to the CDW puddles, and (b) In the region of domain walls between CDW islands. We present the results of these two traces of LDOS in Fig.~(\ref{fig:TDOS}b) for $V=0.8$, a disorder strength that marks the onset of sSC.

The LDOS in the regions of domain wall features an enhanced weight near the Fermi energy ($\omega=0$) in $N(\omega)$. In contrast, the LDOS shows clear gap on the CDW puddles. The enhancement of the spectral weight near the Fermi energy in regions of domains walls, is beneficial for superconductivity to nucleate there. An identical mechanism has also been proposed recently, based on the differential conductance data from scanning tunneling spectroscopy~\cite{YanSTM} in 1T-\chem{TiSe_2} and for \chem{Ti} intercalation in 1T-\chem{TiSe_2}~\cite{TiSTM}.
 
\begin{figure}
      \includegraphics[width=9.0cm,height=4.2cm]{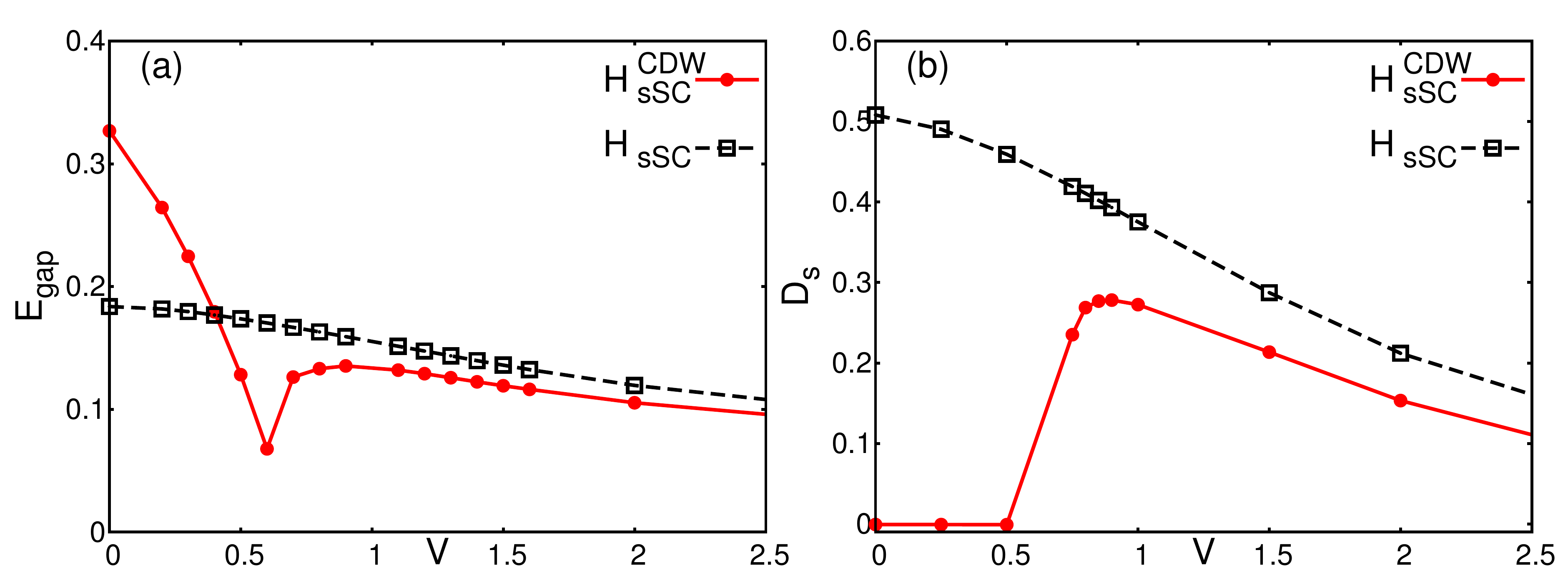}
      \caption{(Color online) Panel \textbf{(a)} shows the evolution energy gap $E_{\rm gap}$ with $V$ for the two models, $\mathcal{H}^{\rm CDW}_{\rm sSC}$ and $\mathcal{H}_{\rm sSC}$, on systems of size $N=54\times54$. Here we track $E_{\rm gap}$ from the  lowest (positive) BdG eigenvalue. The two models show significantly different $E_{\rm gap}$ below $V_1$, whereas the behaviors are quite similar beyond $V_2$. The sharp dip represents the creation of states near the gap edge upon the destruction of global CDW ordering. Panel \textbf{(b)} presents the superfluid stiffness, $D_s$ as a function of $V$ for the two models. Calculation of $D_s$ is carried out on a smaller system ($N=30\times 30$) because of the high numerical demand of such calculations.}  
      \label{fig:egap_ds}
\end{figure}

We have also studied the disorder dependence of the gap in the single particle DOS, which we denote as, $E_{\rm gap}$, in Fig.~(\ref{fig:egap_ds}a). This is obtained by tracking the lowest BdG eigenvalue for a given disorder configuration and then disorder averaging it over many independent configurations. The initial sharp fall of $E_{\rm gap}$ calculated from $\mathcal{H}^{\rm CDW}_{\rm sSC}$ model establishes the destruction of global CDW ordering near $V_1$. The fall arises because of quick filling of states in the mid gap region.

However, further increase of $V$ enhances $E_{\rm gap}$ to a value consistent with that of a disordered superconductor before its downturn at larger values of disorder. This non-monotonic behavior of $E_{\rm gap}$ is very similar to that of the peak height as a function of the disorder strength. To confirm that the final decay of $E_{\rm gap}$ arises from disordered sSC we plotted $E_{\rm gap}$ obtained from $\mathcal{H}_{\rm sSC}$ of Eq.~(\ref{sSCmodel}). $E_{\rm gap}$ in the single particle DOS obtained from the two calculations become almost same \footnote{Since there is no self-consistency for local density in $\mathcal{H}_{\rm sSC}$ model it has a different energy scale for $\mu$ from $\mathcal{H}^{\rm CDW}_{\rm sSC}$ and hence shifted values for observables.} for $V>1.5$ confirming the irrelevance of short ranged CDW fluctuations at those $V$.

The results shown in Fig.~(\ref{fig:TDOS}) are gathered with  enhanced resolution by extending calculations on a much bigger effective system, consisting of $12 \times 12$ identical unit cells each of which is of size $N=54\times54$. Such extension, termed repeated zone scheme~\cite{Kalinsky} is standard and takes advantage of Bloch's theorem for periodically repeated systems.

\subsection{Superfluid Density}
The other defining characteristics of a superconductor is its perfect diamagnetism which leads to Meissner effect. This is because a superconductor develops a stiffness against external magnetic field. This is quantified by superfluid stiffness, and expressed within linear response theory using Kubo formula, as: 
\begin{equation}
D_s=\langle -K_{\rm x} \rangle +\Lambda_{xx}(q_x=0,q_y \rightarrow 0, \omega=0)
\end{equation}
where $K_{\rm x}$ is the kinetic energy along $x$-direction, and $\Lambda_{xx}$ is the long-wavelength limit of (static) paramagnetic current-current correlation~\cite{SWZ}.

We have evaluated $D_s$ for various $V$ as shown in Fig.~(\ref{fig:egap_ds}b). The result from $\mathcal{H}^{\rm CDW}_{\rm sSC}$ at half-filling shows no superfluidity in the system until $V \sim 0.5$, because in this region the subdominant sSC order loses to CDW as the chosen ground state. However, $D_s$ picks up sharply at $V \sim V_1$, a disorder strength where the global CDW order crashes and local superconductivity emerges on the CDW domain walls. Finally, superconducting correlations diminish with increasing $V$ due to localization effects. In fact, the disorder dependence of $D_s$ shows a very close parallel to that of $\Delta_{\rm OP}$. This is expected in a mean field treatment in the presence of disorder~\cite{GTR,Debmalya0}. For a justified comparison we also show the results of $D_s(V)$ calculated from $\mathcal{H}_{\rm sSC}$. The traces from the two calculations become parallel around $V \sim V_2$. The calculation for superfluid stiffness is carried out for smaller system size: $N=30 \times 30$, as the calculation of $D_s$ is numerically demanding. It is for this same reason, calculation of $D_s$ does not implement the repeated zone scheme~\cite{Kalinsky}.

\section{Discussion and conclusion}  
We reported here a study of the interplay of charge density wave order and s-wave superconductivity in two dimensional disordered media within the framework of inhomogeneous Hartree-Fock-Bogoluibov mean field description. We choose model and parameters for our study such that the ground state of the clean system features a CDW ordered state in which the s-wave superconductivity remains sub-dominant.
Upon introduction of disorder, we find that the nature of the ground state changes in an interesting manner. The global CDW order is rapidly destroyed at a weaker disorder strength $V_1$. This allows subdominant superconductivity to emerge in domain walls which separate regions of coherent charge density modulations. This eventually turns the disordered ground state superconducting. The resulting disordered superconductivity survives up to a much larger disorder strength taking advantage of the Anderson's theorem~\cite{AndTheo}! Thus, the introduction of impurities turns a CDW ground state into a superconductor -- an effect quite in contrast to conventional wisdom. 
Most of our results are consistent with the qualitative findings of recent experiments on transition metal dichalcogenides~\cite{Cava0,Abbamonte,YanSTM,Arpes2HNbSe2,Li2017}. This emphasizes the role of intercalation in introducing stoichiometric disorder in these samples, providing crucial insights into the nature of these complex materials. Beyond making pathways for superconductivity, the weak critical disorder has interesting effects on the destruction of CDW ordering -- we find that the local fluctuations of $\pi$-phase shifted CDW ordering (in the form of patches of size $\xi$) persists for larger strengths of disorder, and modifies the nature of disorder superconductivity up to a larger disorder $V_2$ ($< V_1$). Appealingly, commensurate charge modulations separated by phase slips are quite ubiquitous when competing with superconductivity and have recently been observed in underdoped BSCCO, a high temperature cuprate superconductor.~\cite{tJStripe,Mesaros}. Those materials are believed to be intrinsically disordered due to the presence of out-of-plane dopants, a situation somewhat similar to \chem{Cu}-intercalated transition metal dichalcogenides in question, though the mechanism of SC in these materials is expectedly different from that in the dichalcogenides. It will be interesting to address the interplay between the charge order and the superconductivity in the presence of impurities in these unconventional cuprate superconductors as well.

\section*{Acknowledgement}
AB acknowledges IISER-Kolkata for doctoral fellowship. AG acknowledges the hospitality of the International Centre for Theoretical Sciences (ICTS), TIFR, during a visit for a workshop (Code: ICTS/Prog-cqdiscor/2017/05), where part of this research was carried out.
\appendix
\section{Derivation of `gap equations' for the clean system}
The checkerboard CDW breaks the translational symmetry into two sublattices. Therefore, we transform the Hamiltonian in Eq.~(\ref{Comp}) into $N/2$ sites of A sublattice with creation (annihilation) operator $a^{\dagger}(a)$ and $N/2$ sites of B sublattice with creation (annihilation) operator $b^{\dagger}(b)$. We Fourier transform these operators in momentum space where it becomes block diagonal in each $k$-vector and is given by:
\begin{widetext}
\begin{align}
\mathcal{H}^{\rm CDW}_{\rm sSC} = \sum_{k \in HBZ}
  \begin{matrix}\begin{pmatrix} a^{\dagger}_{k\uparrow} & a_{-k\downarrow} & b^{\dagger}_{k\uparrow} & b_{-k\downarrow}\end{pmatrix}\\\mbox{}\end{matrix}
  \begin{pmatrix} -(\alpha \chi_0+ \tilde{\mu}) & \Delta_A & \gamma_k & 0\\ \Delta_A & (\alpha \chi_0+ \tilde{\mu}) & 0 &-\gamma^*_k \\ \gamma^*_k & 0 & (\alpha \chi_0 - \tilde{\mu}) & \Delta_B \\ 0 & -\gamma^*_k & \Delta_B &  -(\alpha \chi_0 - \tilde{\mu}) \end{pmatrix} 
  \begin{pmatrix} a_{k\uparrow} \\ a^{\dagger}_{-k\downarrow} \\ b_{k\uparrow} \\ b^{\dagger}_{-k\downarrow}\end{pmatrix} + E_0
\end{align}
\end{widetext} 
where the sum over $k$ is taken over half of the First Brillioun Zone of the underlying lattice. The dispersion is given by $\gamma_k=(-t-W\Gamma_0) (\rm cos(k_x) + \rm cos(k_y))$ for the square lattice in question. The pairing amplitude on the A-sublattice is denoted as $\Delta_A=\langle a_{k \downarrow} a_{k \uparrow}\rangle$ and for the B-sublattice as $\Delta_B=\langle b_{k \downarrow} b_{k \uparrow}\rangle$. For homogeneous system under consideration, $\Delta_A \equiv \Delta^{\rm(A)}_{\rm OP}$ and similarly for B-sublattice, as used in Fig.~(\ref{fig:phase_dia}). For simplicity of notation we use $\alpha=(U+Wz)/2$ and $\tilde{\mu}=\mu-((U-Wz) \rho)/2$. 
Also the constant terms in the energy of the mean field Hamiltonian is given by:
\begin{equation}
E_{0} =N \left( \frac{\alpha \chi_0^2}{2} + \frac{1}{2U}(\Delta_A^2+\Delta_B^2)+ Wz\Gamma_0^2\right) 
\end{equation}
We diagonalize the above Hamiltonian, and find the quasi-particle excitation spectrum $\omega_q$. The Free energy can then be written as:
\begin{equation}
\mathcal{F}=E_0-\frac{1}{\beta}\sum_q \ln(1+e^{-\beta \omega_q})
\end{equation} 
where $\beta$ is the inverse temperature. The self-consistency equations for $\Delta_A,\Delta_B,\chi_0$ and $\Gamma_0$ is found by extremizing the free energy with respect to the order parameter. The density equation can be obtained by using $\rho=\left(1-\frac{1}{N} \tfrac{d \mathcal{F}}{d \mu}\right)$. We finally take the $T \rightarrow 0$ limit. The phase diagram for the clean case is obtained by solving the five coupled algebraic equations (one each for $\Delta_A, \Delta_B,\chi_0, \Gamma_0$ and $\rho$) numerically and the results are shown in Fig.~(\ref{fig:phase_dia}).
   
\section{Interplay away from half-filling ($\rho=0.95$) in disordered systems}

 \begin{figure}
      \includegraphics[width=8.5cm,height=10.0cm,keepaspectratio]{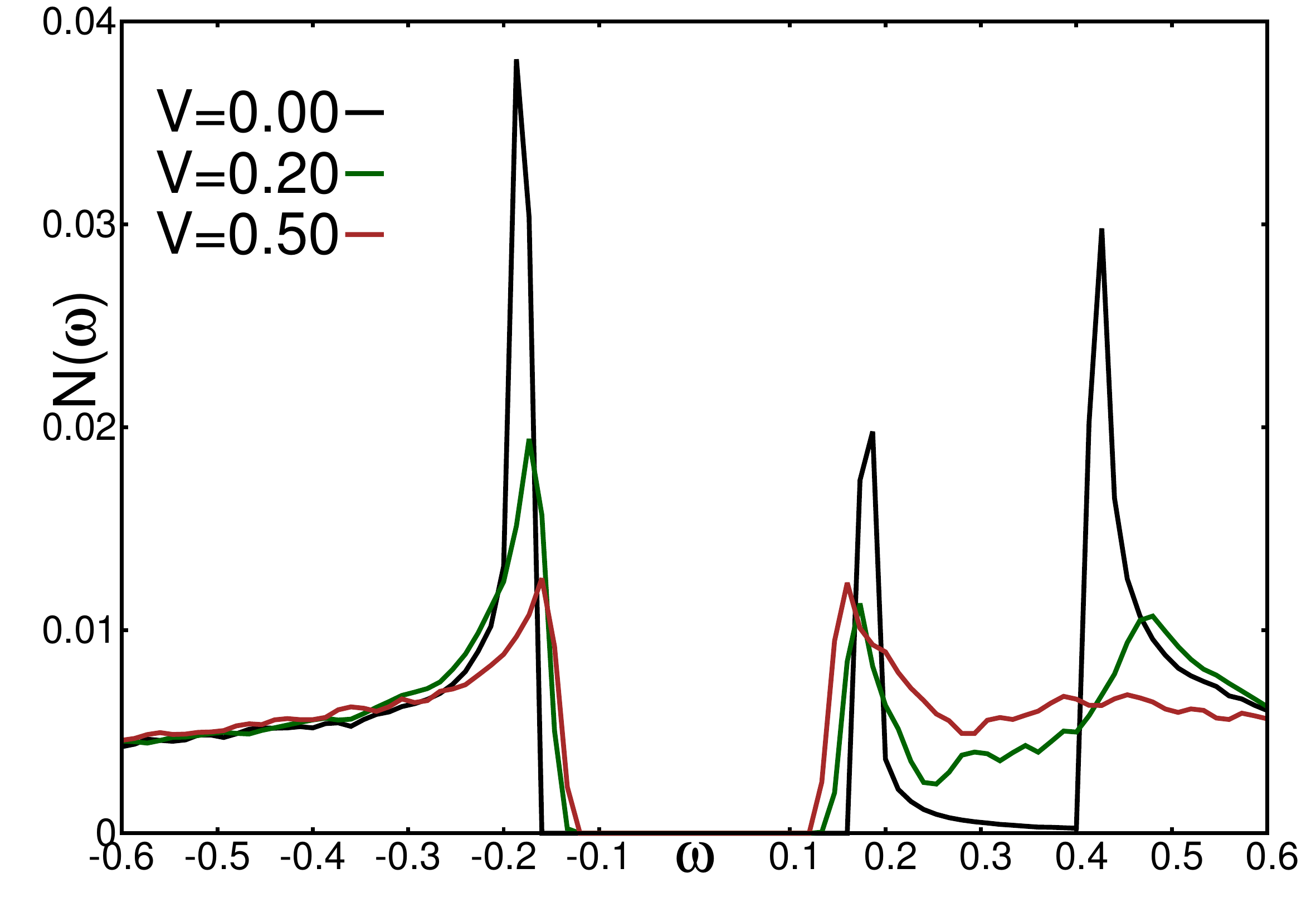}
      \caption{(Color online) The density of states (DOS) averaged over all sites and over disorder configurations for different values of $V$ are shown for $\rho=0.95$. This should be contrasted with Fig.~(\ref{fig:TDOS}) in the main text, which represented half-filling, i.e., $\rho=1$. The traces in the current figure shows a two-gap structure, most prominent for $V=0$, arising individually from sSC and CDW orders. With the increase of $V$ global CDW depletes quickly, and the traces at large $V$ becomes indistinguishable from those for a half-filled system.}  
      \label{fig:TDOS_n95}
\end{figure}  

The interplay of CDW and sSC orders in the presence of disorder for a system with $\rho=0.95$ is broadly similar to the results discussed in the main paper for a system kept at half-filling ($\rho=1$). However, the signatures of charge modulations are weaker, as expected. For example, the evolution of the $\bf{q}=(\pi,\pi)$ peak of $S(\bf{q})$ with disorder is exactly similar to the case with $\rho=1$, except for the fact that the intensity of the peak is much weaker, for corresponding disorder values.
   \begin{figure}
      \includegraphics[width=10.0cm,height=12.0cm,keepaspectratio]{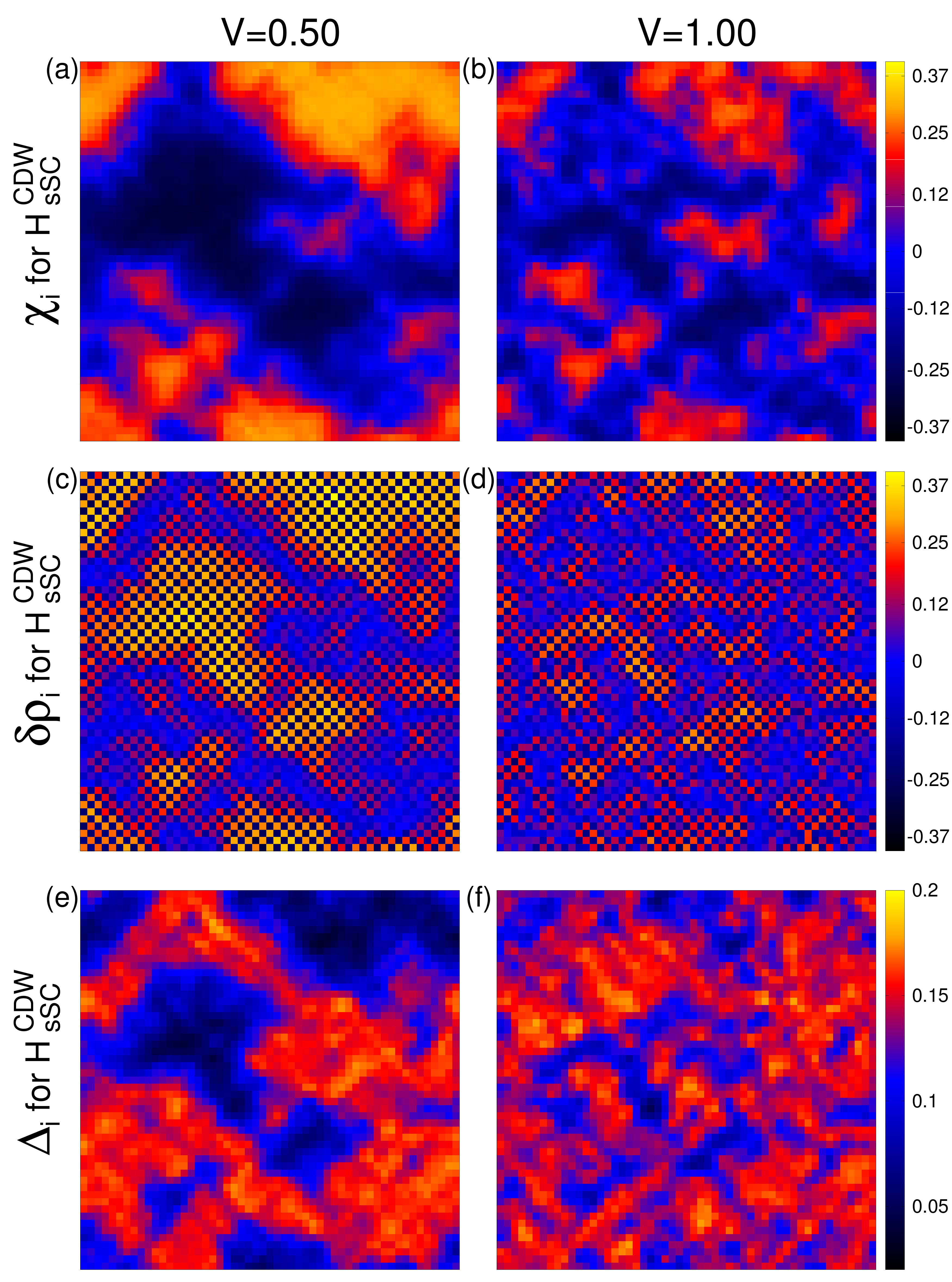}
      \caption{
(Color Online) Spatial profile of order parameters for $\rho=0.95$, in model $\mathcal{H}^{\rm CDW}_{\rm sSC}$:\\
Panel \textbf{(a)} shows the spatial profile of $\chi_i$ for $V=0.5$, while, panel \textbf{(b)} presents silmilar data for $V=1.0$. Panels \textbf{(c)} and \textbf{(d)} show the spatial profiles of the density modulation for $V=0.5$ and $V=1.0$ respectively. Finally, panels \textbf{(e)} and \textbf{(f)} present the superconducting pairing amplitudes for $V=0.5$ and $V=1.0$ respectively. The spatial separation of the two independent orders is apparent from these results similar to what was found in Fig.~(\ref{fig:LOP}) at half-filling.}  
      \label{fig:LOP_n95}
\end{figure}

A major difference in the results from the system at $\rho=0.95$ compared to that of the half-filled system occurs in the evolution of the average DOS with disorder. Away from the half-filling the DOS has a characteristic two gap structure (being most prominent in the clean system) as shown in Fig.~(\ref{fig:TDOS_n95}). The two gaps arise from the two individual orders as the ground state features both the orders at $\rho=0.95$. Both of these broken symmetry orders open up their own gap, but at different energies. The gap associated with the sSC order appears at $\omega-\mu=0$, whereas, the CDW order opens up its characteristic gap at the chemical potential, $\mu$ (Note that $\mu=0$ for a half-filled system). As $V$ increases, the CDW gap (at $\omega = \mu \approx 0.4$) fills up quickly with the fast weakening of coherent charge modulation, and ultimately we are left with a single gap in the DOS, by $V \sim V_1$, where global CDW is lost. The coherence peak at the gap edge from sSC order initially increases upon the destruction of global CDW ordering, however, the sSC coherence peak finally reduces for larger $V$. 

We also show the spatial variation of the order parameter and density fluctuation for $\rho=0.95$ in Fig.~(\ref{fig:LOP_n95}). Here we show result for two representative disorder strengths. In the left panels we show results for a disorder $V=0.5<V_1$, and for $V=1.0>V_1$ on the right panels. 
The top panels show charge modulation amplitude $\chi_i$. We see that even for $V<V_1$, CDW order is broken in positive and negative patches. However, positive patches are much larger than the negative ones and hence the overall CDW order survives. Fragmentation of CDW order into puddles are found for $V$ as low as $0.2$ for this particular case with $\rho=0.95$. At larger $V$, we find CDW puddles to shrink in size, as seen in Fig.~(\ref{fig:LOP_n95}b). This is similar to the results in the main text for a half-filled system.
The spatial separation of the regions of these two ordering is evident by comparing the modulation of density, as shown Fig.~(\ref{fig:LOP_n95}c, \ref{fig:LOP_n95}d). The spatial profiles of the sSC pairing amplitude are shown in Fig.~(\ref{fig:LOP_n95}e, \ref{fig:LOP_n95}f). We also notice the sSC pairing is uniform in a blob, unlike in clean system where pairing amplitude modulated in response to the density modulation, as seen in Fig.~(\ref{fig:phase_dia}b). With further increase of $V$, the CDW puddles become very small, so that they have no discernible effect on the disordered sSC.

\begin{figure}
      \includegraphics[width=8.0cm,height=8.0cm,keepaspectratio]{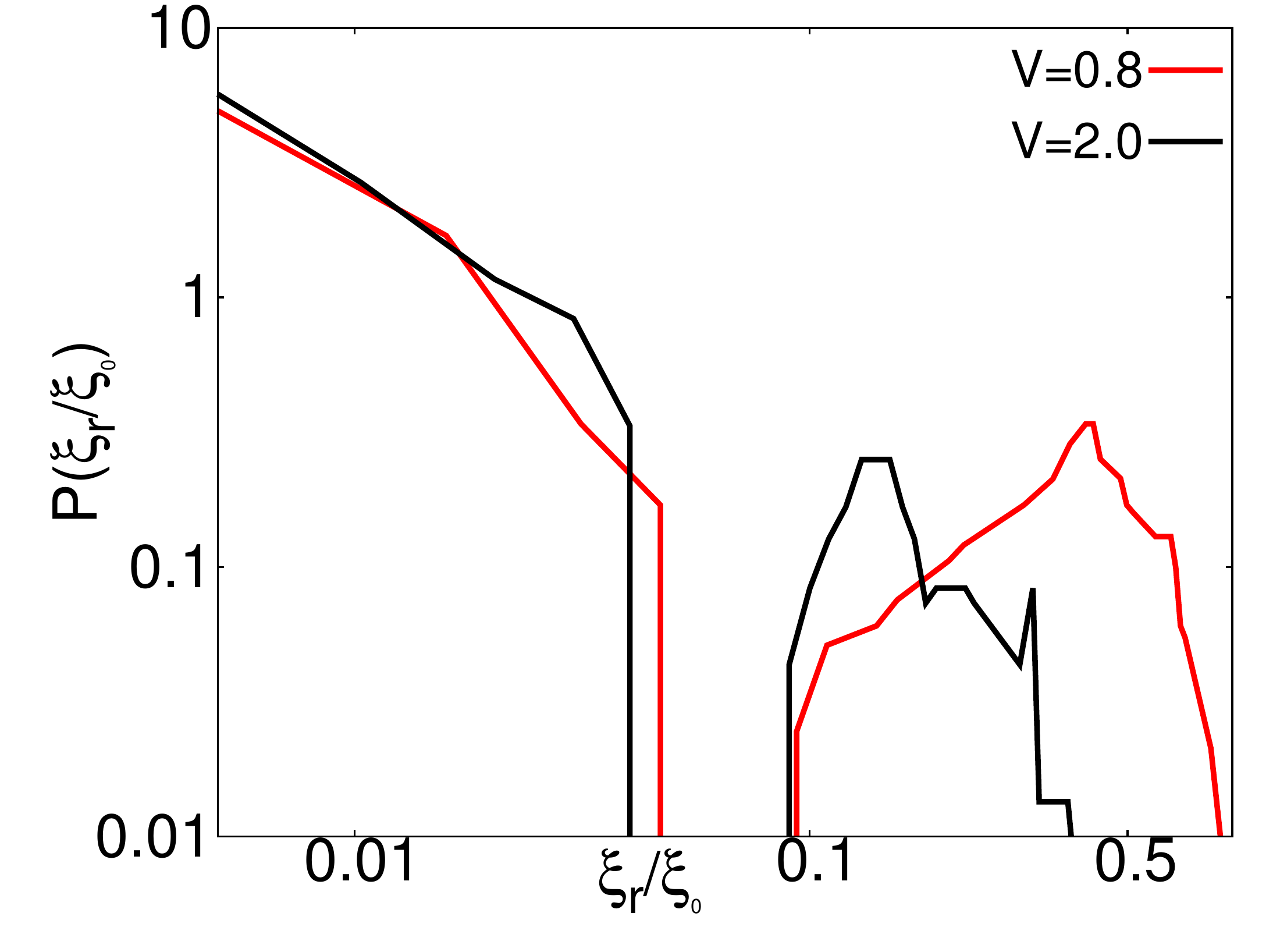}
      \caption{(Color online) Analysis of the size distribution of the typical CDW cluster is illustrated by showing the distribution of $\xi_r/\xi_{0}$ on log-log scale. Here, $\xi_{0}\equiv \xi_{V=0}$. This distribution is found bimodal for all $V$. A large number of small clusters (or better called `grains') constitute the smaller-$\xi_r/\xi_{0}$ part of the distribution $P(\xi_r/\xi_{0})$. Whereas, there are few large clusters, which produces the second peak of the distribution for larger values of its argument. It is this second part of this distribution which identify the size of the typical clusters. The average of this second part of the distribution defines the length scale of typical clusters.} 
      \label{fig:appC}
\end{figure}

\section{Extracting CDW domain size from real space cluster analysis}

We begin by noting that the CDW amplitude $\chi_i$ takes both positive and negative values. In addition, we find that the positive and negative amplitude of staggered field of modulation remain spatially clustered (as seen in Fig.~(\ref{fig:LOP}a, \ref{fig:LOP}b)). To obtain the domain size of these clusters, we first define the boundary of these domains. To this end, we start from a particular site and note the sign of $\chi_i$ on that site. We consider the nearest neighboring sites to be a part of that same cluster if the sign of $\chi_i$ for the neighboring site is same as that of the given site. We keep repeating the aforementioned step for all the sites in that cluster, until: (a) All possible nearest neighboring sites are of opposite sign, in which case we reach the boundary of that cluster. (b) When all the sites of the system are visited. For our two-dimensional system, the square root of the number of sites belonging to a cluster defines the typical size of the domain or cluster under consideration.

We present the distribution of the normalized domain size ($\xi_r/\xi_0$), where $\xi_0 \equiv \xi(V=0)$, in Fig.~(\ref{fig:appC}) using log-log scale, for two values of disorder strength: $V=0.8$ and $V=2.0$. The striking feature of this distribution is its bimodal nature, which we address below. We find that for all $V$, there are large number of domains of very small sizes (consisting of a few lattice spacing, $\lesssim 5\%$ of the system-size). These ultra-small `grains' of charge modulation have little significance for the existence of the peak in $S(\bf{q})$. In fact, the positive and negative grains of such rapid charge modulation washes out any peak in $S(\bf{q})$. It is these small grains of varying size produce the initial part of the bimodal structure of $P(\xi_r/\xi_0)$, containing a short and monotonically decaying tail. Instead, the major contribution to the peak at $S(\pi,\pi)$ comes from the occasional large domains which span through a macroscopic region of the system. These clusters contribute to the second part of the bimodal structure of $P(\xi_r/\xi_0)$. Therefore, to obtain the typical cluster size, and to compare this with the one obtained from the width of the $S(\pi,\pi)$ peak ($\xi$), we need to consider only these macroscopic clusters.

Thus we use only the large-argument part of $P(\xi_r/\xi_0)$ to obtain the 
typical cluster-size, $\xi_r$, which we plotted in Fig.~({\ref{fig:xi}}). The reduction of the average CDW domain size with the increase of $V$ is also explained by Fig.~(\ref{fig:appC}).

\bibliography{References.bib}
\end{document}